\documentclass[11pt]{article}
\usepackage{fullpage}
\pdfoutput=1
\usepackage{latexsym}
\usepackage{amssymb}
\usepackage{epsfig,amsmath,graphics}
\usepackage{listings}
\usepackage{color}
\numberwithin{equation}{section}

\definecolor{colorLink}{rgb}{0.6,0,0}
\definecolor{colorCite}{rgb}{0,.6,0}
\definecolor{colorURL}{rgb}{0,0.6,0.0}
\usepackage{dcolumn}
\usepackage{slashed}
\usepackage{cancel}
\usepackage{comment,latexsym}
\usepackage{bm}
\usepackage{verbatim}
\usepackage{tabularx}
\usepackage{dcolumn}
\usepackage[colorlinks=true,linktocpage=true,linkcolor=black,citecolor=colorCite,urlcolor=colorURL]{hyperref}
\usepackage{setspace}
\usepackage{xspace}
\usepackage{multirow}
\usepackage{titlesec}
\usepackage[bbgreekl]{mathbbol}
\usepackage{booktabs}
\usepackage{makecell}
\usepackage{array}
\usepackage{hyperref}
\usepackage{bbm}
\usepackage{bm}
\usepackage{amsfonts}
\usepackage{mathrsfs}
\usepackage{mathtools}
\usepackage{bbold}
\usepackage{amsmath,amssymb}
\usepackage{epsfig}
\usepackage{graphicx}               
\usepackage{url}
\usepackage{hyperref}
\usepackage{float}
\usepackage{pstricks}
\usepackage{color}
\usepackage{multirow}
\usepackage{slashed}
\usepackage{hepunits}
\usepackage{cite}

\def\mysection#1{{\bf #1.} }

\newcommand{\lsim}{\!\mathrel{\hbox{\rlap{\lower.55ex \hbox{$\sim$}} \kern-.34em \raise.4ex \hbox{$<$}}}}
\newcommand{\gsim}{\!\mathrel{\hbox{\rlap{\lower.55ex \hbox{$\sim$}} \kern-.34em \raise.4ex \hbox{$>$}}}}
\newcommand{\Tr}{{\rm Tr}}

\def\eg{{\it e.g.}}

\def\be{\begin{equation}}
\def\ee{\end{equation}}
\def\beq{\begin{equation}}
\def\eeq{\end{equation}}
\def\bea{\begin{eqnarray}}
\def\eea{\end{eqnarray}}
\def\bit{\begin{itemize}}
\def\eit{\end{itemize}}
\newcommand{\eref}[1]{(\ref{#1})}
\newcommand{\Eref}[1]{Eq.~(\ref{#1})}

\newcommand{\Erefs}[2]{Eqs.~(\ref{#1}) and~(\ref{#2})}

\def\to{\rightarrow}
\newcommand{\vecK}{\mathbf{K}}
\newcommand{\vecG}{\mathbf{G}}

\newcommand{\vecl}{{\bm \ell}}
\newcommand{\lt}{\widetilde{ \ell}}
\newcommand{\veclt}{\widetilde{{\bm \ell}}}
\newcommand{\veck}{\mathbf{k}}
\newcommand{\vecq}{\mathbf{q}}

\newcommand{\vmin}{v_{\rm min}}
\newcommand{\ang}{{\rm \AA}}

\titleformat{\section}{\center\normalfont\fontsize{14}{15}\bfseries}{\thesection.}{1em}{}
\titleformat{\subsubsection}{\center\normalfont\fontsize{12}{15}}{\thesubsubsection.}{1em}{}

\begin{document}
\begin{spacing}{1.3}
\hfill PUPT-2535
\begin{center}

{\Large \bf
Detection of sub-MeV Dark Matter with \\Three-Dimensional Dirac Materials} \\
\vspace{0.5cm}

 { Yonit Hochberg${}^{1,2}$, Yonatan Kahn${}^{3}$, Mariangela Lisanti${}^{3}$, Kathryn M. Zurek${}^{4,5}$, Adolfo G. Grushin${}^{6,7}$, Roni Ilan${}^{8}$, Sin\'{e}ad M. Griffin${}^{6,9}$, Zhen-Fei Liu${}^{6,9}$, Sophie F. Weber${}^{6,9}$, and Jeffrey B. Neaton${}^{6,9,10,11}$}

 \vspace*{.4cm}
${}^1$ {\it Department of Physics, LEPP, Cornell University, Ithaca NY 14853, USA} \\
${}^2${\it Racah Institute of Physics, Hebrew University of Jerusalem, Jerusalem 91904, Israel} \\
${}^3$ {\it Department of Physics, Princeton University, Princeton, NJ 08544, USA}\\
${}^4$ {\it Ernest Orlando Lawrence Berkeley National Laboratory, \\University of California, Berkeley, CA 94720, USA} \\
${}^5$ {\it Berkeley Center for Theoretical Physics, University of California, Berkeley, CA 94720, USA}\\
${}^6$ {\it Department of Physics, University of California, Berkeley, CA 94720, USA}\\
${}^7$ {\it Institut Ne{\'e}l, CNRS and Universit{\'e} Grenoble Alpes, F-38042 Grenoble, France}\\
${}^8$ {\it Raymond and Beverly Sackler School of Physics and Astronomy, \\Tel-Aviv University, Tel-Aviv 69978, Israel}\\
${}^9$ {\it Molecular Foundry, Lawrence Berkeley National Laboratory, Berkeley, CA 94720, USA}\\
${}^{10}$ {\it Kavli Energy NanoScience Institute at Berkeley, Berkeley, CA 94720, USA}\\
${}^{11}$ {\it Materials Sciences Division, Lawrence Berkeley National Laboratory, \\Berkeley, California 94720, USA}\\
\vspace*{.5cm}

\vspace{-0.2cm}
\end{center}

\begin{abstract}\noindent
We propose the use of three-dimensional Dirac materials as targets for direct detection of sub-MeV dark matter. Dirac materials are characterized by a linear dispersion for low-energy electronic excitations, with a small band gap of $\mathcal{O}(\meV)$ if lattice symmetries are broken.
Dark matter at the keV scale carrying kinetic energy as small as a few meV can scatter and excite an electron across the gap.
Alternatively, bosonic dark matter as light as a few meV can be absorbed
by the electrons in the target. We develop the formalism for dark matter scattering and absorption in Dirac materials and calculate the experimental reach of these target materials.
We find that Dirac materials can play a crucial role in detecting dark matter in the keV to MeV mass range that scatters with electrons via a kinetically mixed dark photon, as the dark photon does not develop an in-medium effective mass. The same target materials provide excellent sensitivity to absorption of light bosonic dark matter in the meV to hundreds of meV mass range, superior to all other existing proposals when the dark matter is a kinetically mixed dark photon.
\end{abstract}

\vspace*{3mm}


\tableofcontents

\newpage
\section{Introduction}
\label{sec:introduction}

The search for sub-GeV dark matter (DM) is a growing frontier in direct detection experiments.  This program is driven by a theoretical revolution revealing a wide and growing range of models for light DM.  In these scenarios, the DM typically resides in a hidden sector with either strongly or weakly interacting dynamics~\cite{Boehm:2003hm,Strassler:2006im,Hooper:2008im,Pospelov:2007mp,ArkaniHamed:2008qp,Feng:2008ya,Foadi:2008qv,Ryttov:2008xe,Alves:2009nf,Frandsen:2009mi,Kribs:2009fy,Lisanti:2009am,Mardon:2009gw,Morrissey:2009ur,Alves:2010dd,Belyaev:2010kp,Cohen:2010kn,Lewis:2011zb,Hietanen:2012qd,Blinov:2012hq,Cline:2013zca,Bai:2013xga,Appelquist:2014jch,Detmold:2014qqa,Detmold:2014kba,Appelquist:2015zfa,Appelquist:2015yfa,Mitridate:2017oky}. There are many ways to fix the observed DM abundance in these theories, including asymmetric DM~\cite{Kaplan:2009ag,Petraki:2013wwa,Zurek:2013wia}, freeze-in~\cite{Hall:2009bx,Bernal:2017kxu}, strong dynamics~\cite{Hochberg:2014dra,Hochberg:2014kqa,Harigaya:2016rwr}, kinematic thresholds~\cite{DAgnolo:2015ujb}, and various non-standard thermal histories~\cite{Kuflik:2015isi,Pappadopulo:2016pkp,Dror:2016rxc,Kuflik:2017iqs,DAgnolo:2017dbv,Berlin:2017ftj}, to name a few.  The breadth of possible scenarios has stimulated a rethinking of the ideal experimental targets for discovery.

Directly detecting DM relies on observing the effects of its interactions with an experimental target, either through scattering or absorption in the material.  In both cases, sufficient energy must be deposited to observe the interaction; this becomes increasingly challenging as the DM mass is reduced.  The current suite of direct detection experiments focuses on the weakly-interacting massive particle (WIMP), where the DM mass is typically above $\sim 10 ~\mbox{GeV}$.  These experiments search for nuclei that recoil after a collision with a DM particle.  Since the energy deposited in an elastic scattering process is $q^2/2 m_T$, where $q$ is the momentum transfer and $m_T$ is the mass of the target, it often becomes more effective to search for energy deposition on electron targets when DM is less massive than a nucleus.  Condensed-matter systems are sensitive to scattering events where the DM carries comparable kinetic energy to the electron excitation energy. For many such systems, including semiconductors \cite{Essig:2011nj,Essig:2012yx,Graham:2012su}, graphene \cite{Hochberg:2016ntt}, scintillators \cite{Derenzo:2016fse}, molecules \cite{Essig:2016crl}, and crystal lattices \cite{Budnik:2017sbu}, these energies are eV-scale. This is optimal for detecting DM $\chi$ with mass $m_\chi \gtrsim \text{ MeV}$, where the kinetic energy is $m_\chi v_\chi^2/2$ with $v_\chi \sim 10^{-3}$, the virial velocity of DM in the Galaxy.\footnote{Throughout this paper, we use natural units with $\hbar = c = 1$; all velocities are expressed in units of $c$ and all distances in units of momentum.}  If instead, $\chi$ is a boson with mass $\gtrsim$~eV, it can be detected via absorption on an electron in these same systems~\cite{Hochberg:2016sqx,Bloch:2016sjj}.

Extending experimental sensitivity to scattering or absorption of even lower mass DM carries many challenges.  For example, fermionic DM is consistent with all astrophysical observations when its mass is greater than a few keV, but to reach these mass scales, one must find a material where the few meV of energy it deposits in scattering can lead to observable signatures.  Superconducting targets offer one promising option~\cite{Hochberg:2015pha,Hochberg:2015fth,Hochberg:2016ajh}.  These ultra-pure materials, with a small ($\sim \meV$) gap and a large Fermi velocity, are sensitive to DM scatters in the keV--MeV mass range or to meV--eV DM absorption.   Superfluid helium has also been shown to be sensitive to sub-MeV DM, when the DM collision can produce multiple phonons~\cite{Schutz:2016tid,Knapen:2016cue}. Neither superconductors nor superfluid helium, however, have optimal sensitivity to dark photons \cite{Holdom:1985ag,Okun:1982xi}, which can serve either as the mediator for DM-electron scattering processes or as the DM itself which is absorbed.  In the case of superconductors, the dark photon takes on a large effective mass in the medium, suppressing the DM interaction rate.  For helium, the leading interaction is through the polarizability of the atom, which is small.

\begin{figure}[t] 
   \centering
   \includegraphics[width=0.35\textwidth]{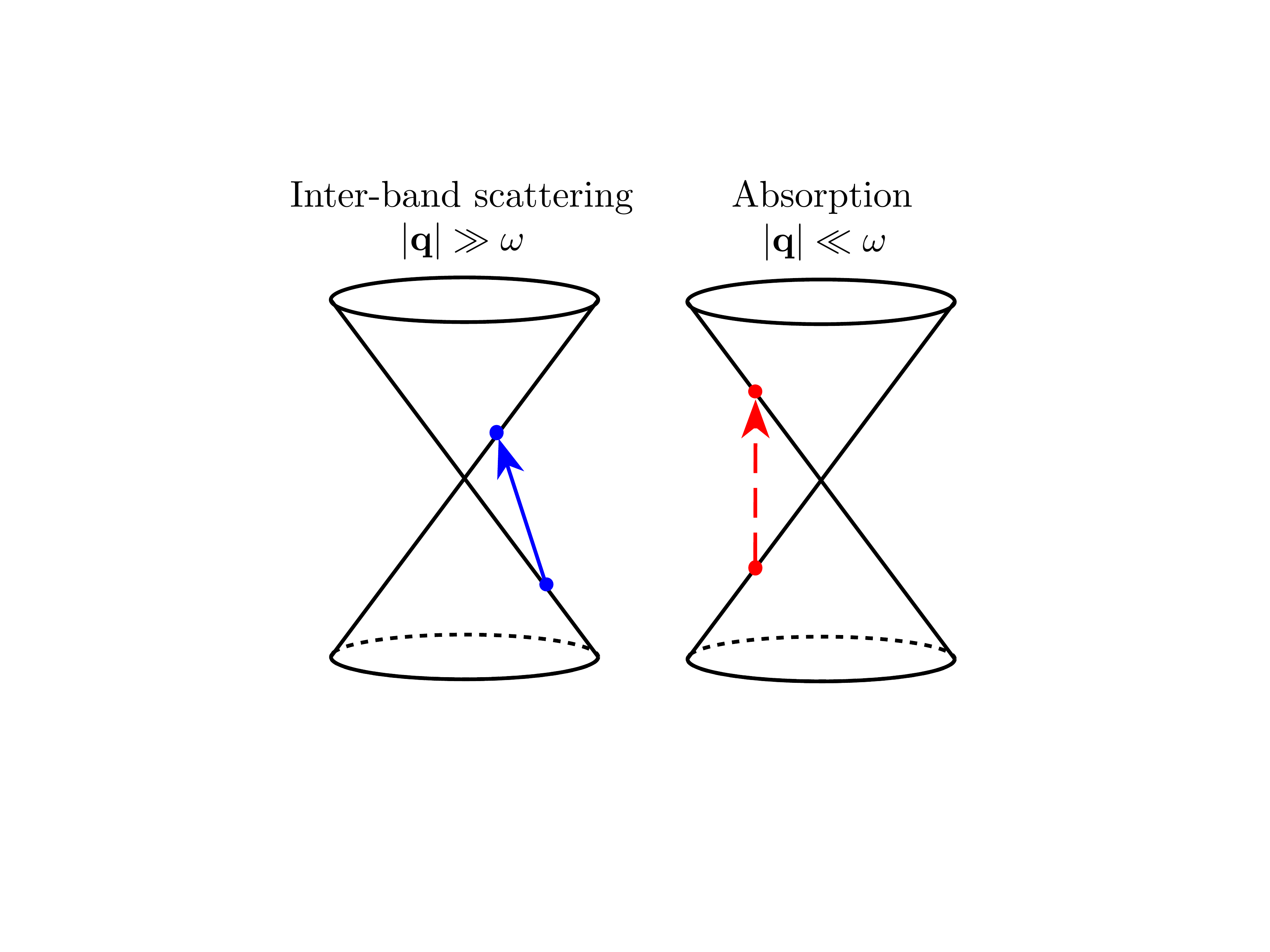}
   \caption{Cartoon of the two dark matter-initiated processes in Dirac materials that we consider in this paper: inter-band (valence to conduction) scattering (\textit{left}) and absorption by valence-band electrons (\textit{right}).}
   \label{fig:schematic}
\end{figure}

In this paper, we propose Dirac materials as a new class of electron targets for DM scattering or absorption. We define Dirac materials as three-dimensional (3D) bulk substances whose low-energy electronic excitations are characterized by a Dirac Hamiltonian~\cite{HQ13,VV14,AMV17},
\be
H_{\vecl} =   \left ( \begin{matrix}
      0 & v_F \vecl \cdot \bm{\sigma} - i \Delta \\
      v_F \vecl \cdot \bm{\sigma} + i \Delta & 0 \\
   \end{matrix} \right), \qquad E^{\pm}_{\vecl} = \pm \sqrt{v_F^2 \vecl^2 + \Delta^2}.
 \label{eq:DiracHam}
\ee
Here, $\vecl$ is a lattice momentum measured from the location of the point of the Dirac cone (\emph{e.g.}, the Dirac point) in reciprocal space, $\Delta$ is analogous to the mass term in the Dirac equation giving rise to a band gap $2\Delta$, the Fermi velocity $v_F$ plays the role of the speed of light $c$, and the positive and negative dispersion relations correspond to the conduction and valence bands, respectively.\footnote{Real materials typically have anisotropic Fermi velocities, but this complication does not affect the thrust of our arguments; we treat this case in Appendices~\ref{app:Overlap} and~\ref{app:aniso}.} The desired signal is a DM-induced inter-band transition from the valence to the conduction band, where for DM scattering the momentum transfer $|\vecq|$ is typically much larger than the energy deposit $\omega$, with the opposite being true for absorption of non-relativistic DM. A cartoon of these two processes is illustrated in Fig.~\ref{fig:schematic}. As we will show, the dynamics of the photon interacting with Dirac fermions mimic those of ordinary relativistic QED: the Ward identity keeps the photon massless in a Dirac material, leading to excellent detection reach in models of DM involving dark photons.

When $\Delta = 0$, the low-energy degrees of freedom in a Dirac material correspond to two Weyl fermions of opposite chiralities. Materials with this feature are classified as either Dirac or Weyl semimetals and are regarded as the 3D analogues of graphene. In Dirac semimetals, both Weyl fermions occur at the same point in momentum space, but are decoupled
due to an additional crystalline symmetry which imposes $\Delta = 0$. Examples of Dirac semimetals include Na$_3$Bi~\cite{WSC12,Liu864} and Cd$_3$Ar$_2$ \cite{WWH13,LJZ14,2016PhRvB..94h5121J}. Allowing the two Weyl fermions to couple, for example by applying strain to a Dirac semimetal or tuning a topological insulator close to the semimetal critical point~\cite{Xu560}, can lead to a finite $\Delta \neq 0$ that is typically small, $2\Delta \sim \meV$. Such a gap can suppress thermal inter-band transitions, which is crucial for making detection of meV-scale DM-induced excitations feasible.\footnote{In Weyl semimetals, the two Weyl fermions are generically located at different points in momentum space and thus are decoupled at low energies~\cite{WFC15,HXB15,Xu613,LWF15}, making it difficult to open a gap. As the gap is necessary to control thermal noise in our proposal, we do not consider these materials further in this paper.} While our analysis is completely general, we propose ZrTe$_5$ as a realistic target Dirac material. ZrTe$_5$ has been synthesized experimentally, and in this work we compute its band structure from first principles, finding in particular that its small Fermi velocities and tunable Fermi level, which can be located inside the gap, make it especially suitable for a dark matter search.

This paper is organized as follows.  Section~\ref{sec:formalism} presents the benchmark dark photon model, and then introduces the formalism for describing in-medium effects in Dirac materials.  This formalism is used in Sections~\ref{sec:scattering} and~\ref{sec:absorption} to calculate the DM scattering rate mediated by a dark photon and the dark photon absorption rate in Dirac materials, respectively.  For both cases, we present sensitivity projections for couplings to electrons, comparing them to other proposals for sub-MeV dark matter detection.   We conclude in Section~\ref{sec:conclusions} with a brief discussion of experimental considerations. The four Appendices describe the derivation of the transition form factor for a generic Dirac material, the generalization of the scattering rate to anisotropic semimetals, the scattering and absorption reach for models other than the light kinetically mixed dark photon, and the density functional theory (DFT) calculations used to derive the band structure of ZrTe$_5$.
\section{\textbf{Dark Matter Interactions with In-Medium Effects}}
\label{sec:formalism}

Our discussion of sub-MeV DM is focused on the benchmark model of the kinetically mixed dark photon.  Specifically, we consider a new $U(1)_D$ gauge boson that mixes with the ordinary photon:
\beq
{\cal L} = -\frac{1}{4} F_{\mu \nu} F^{\mu \nu} - \frac{1}{4} F'_{\mu \nu} F'^{\mu \nu} - \frac{\varepsilon}{2} F_{\mu \nu} F'^{\mu \nu} + e J_\text{EM}^\mu A_\mu + g_D J^\mu_\text{DM} A'_\mu + \frac{m_{A'}^2}{2} A'^{\mu} A'_{\mu} \, .
\label{eq:kinmixing}
\eeq
Here, $F_{\mu \nu}$ ($F'_{\mu \nu}$) is the ordinary (dark) electromagnetic field strength, $\varepsilon$ is the kinetic mixing parameter, and $J^\mu_\text{EM (DM)}$ is the electromagnetic (dark) current, which couples to the (dark) photon with strength $e$ ($g_D$).\footnote{In this paper, we follow high-energy physics conventions and use Heaviside-Lorentz units for electromagnetism, where $e = \sqrt{4\pi\alpha_{\rm EM}} \simeq \sqrt{4\pi/137}$.} We assume that the new dark photon field $A'_\mu$ acquires a mass $m_{A'}$ either through a dark Higgs or Stueckelberg mechanism.  The propagating dark photon $\widetilde{A}'_\mu$ in the mass basis can be identified by diagonalizing the kinetic terms in Eq.~\eqref{eq:kinmixing}, and can serve as either the DM itself or as a mediator of the interactions between the Standard Model and the DM which comprises the dark current $J^\mu_\text{DM}$.

Due to the induced coupling of the dark photon to the electromagnetic field strength, dark photon interactions are modified in an optically responsive medium. The effects of the medium on the dark photon coupling can be derived by considering the effects of the medium on an ordinary photon, where the propagator is modified via its interactions with the medium. One finds \cite{Hochberg:2015fth} that the transverse and longitudinal dark photon fields ${\tilde{A'}_\mu}^{T,L}$ interact with the electromagnetic current with reduced coupling:
\beq
{\cal L} \supset \varepsilon e \frac{q^2}{q^2-\Pi_{T,L}} \tilde{A'}_\mu^{T,L} J^\mu_{\rm EM}\,.
\label{eq:massbasis}
\eeq
Here, $\Pi_{T,L}$ are the transverse and longitudinal components of the in-medium polarization tensor,
$
\Pi^{\mu\nu} = \Pi_T \sum_{i=1,2} \epsilon_i^{T\mu} \epsilon_i^{T*\nu} + \Pi_L \epsilon^{L\mu}\epsilon^{L\nu}\,,
$
with
$
\epsilon^L  =  \frac{1}{\sqrt{q^2}}(|{\bf q}|,\omega \frac{{\bf q}}{|\bf q|})$ and
$ \epsilon^T_{1,2}  = \frac{1}{\sqrt{2}} (0,1,\pm i,0)\,.
$
As a result of \Eref{eq:massbasis}, dark photon interactions inside a medium depend on the electromagnetic response of the medium, parameterized by $\Pi_{T,L}$ (see detailed discussion in Ref.~\cite{Hochberg:2015fth}).  In this section, we describe the behavior of an ordinary photon in an optically responsive medium.  We review the optical properties of Dirac materials in Section~\ref{ssec:opticalSM} and compare the results to that of metals in Section~\ref{ssec:metalcomparison}.  We will use these results to model dark photon scattering and absorption processes in later sections of the paper.

\subsection{Optical Properties of Dirac Materials}
\label{ssec:opticalSM}

In Lorentz gauge, the in-medium photon propagator is written as
\beq
G^{\mu\nu}_\text{\rm med} (q) = \frac{P_{T}^{\mu\nu}}{\Pi_T -q^2}  + \frac{P_{L}^{\mu\nu}}{\Pi_L -q^2} \, ,
\label{eq:GInMedium}
\eeq
where $q = \left(\omega, \mathbf{q} \right)$ is the 4-momentum transfer, $q^2 = \omega^2 - \vecq^2$, and $P_{L,T}$ are longitudinal and transverse projection operators, respectively (see \emph{e.g.}, Ref.~\cite{Schmitt:2014eka} for a complete derivation).  From Eq.~(\ref{eq:GInMedium}), we see that the photon can develop an effective mass in-medium if the real part of $\Pi_{T,L}(q)$ contains terms that do not vanish at $q^2 = 0$.  In general, $\Pi_{T,L}(q)$ may be a complicated function of $q$ with no simple interpretation as an effective mass, but large $\Pi_{T,L}$ will generally suppress electromagnetic interactions. The imaginary parts of $\Pi_{T,L}$ determine the probability of photon absorption.

The transverse and longitudinal components of the in-medium polarization tensor are linked to the optical response of the medium through the complex permittivity $\epsilon_r$ by
\bea
 \Pi_L = q^2(1-\epsilon_r)   \quad \text{ and } \quad
\Pi_T = \omega^2(1-\epsilon_r)  \,.
\label{PiLPiT}
\eea
In the regime $|q^2| \sim \vecq^2 \gg \omega^2$, which is relevant for DM scattering, $\Pi_L$ dominates over $\Pi_T$. Conversely, in the case of DM absorption where $q^2 \sim \omega^2\gg \vecq^2$, $\Pi_L \simeq \Pi_T$.

For Dirac materials with a band gap, it is simplest to determine the complex permittivity $\epsilon_r$  by borrowing the expression for the one-loop polarization function in massive QED in 3+1 dimensions (see {\it e.g.}, Ref.~\cite{Schwartz}).  In doing so, we substitute $c \to v_F$ and $\alpha_\text{EM} \to \widetilde{\alpha}$, where $v_F$ is the Fermi velocity and $\widetilde{\alpha}$ is the effective fine-structure constant in the medium:
\be\label{eq:alphatilde}
\widetilde{\alpha} = \alpha_\text{EM} \times \frac{g}{\kappa v_F} \, ,
\ee
with $\kappa$ the background dielectric constant, $\alpha_\text{EM} = e^2/4 \pi$, and $g = g_s g_C$ is the product of spin and Dirac cone degeneracy~\cite{throckmorton2015many}.  In the $\overline{MS}$ scheme, to leading order in $\widetilde{\alpha}$, the complex permittivity (at zero temperature and doping) is therefore given by:
\begin{align}
\left(\epsilon_r\right)_{\rm Dirac} = 1 + &\frac{e^2 g}{4\pi^2\kappa v_F}\int_0^1 dx\, \left \{ x(1-x) \ln \left |\frac{(2 v_F \Lambda)^2}{\Delta^2 - x(1-x)(\omega^2 - v_F^2 \vecq^2)}\right| \right \} \nonumber \\
& + i\frac{e^2 g}{24\pi \kappa v_F}\sqrt{1 - \frac{4\Delta^2}{\omega^2 - v_F^2 \vecq^2}}\left (1 + \frac{2\Delta^2}{\omega^2 - v_F^2 \vecq^2}\right)\Theta(\omega^2 - v_F^2 \vecq^2 - 4\Delta^2) \, ,
\label{eq:epsGap}
\end{align}
where $\Lambda$ is a UV cutoff, defined as the momentum distance from the Dirac point at which the dispersion relation deviates from linear.\footnote{Here we are effectively setting the renormalization scale $\tilde{\mu}$ at the cutoff, $\tilde{\mu} = 2 v_F \Lambda$, which is perhaps unusual from a high-energy physics perspective. The unphysical parameter $\tilde{\mu}$ can be removed from physical quantities by matching to a measurement of the electric charge $e$.  In QED, one typically thinks of the electric charge as being defined by a $t$-channel scattering process, \eg\ $e^- + e^- \to e^- + e^-$. However, the inter-band transition in a Dirac material is analogous to pair production, which is an $s$-channel process.  DM scattering in Dirac materials can be described by $\chi + N \to \chi + N + \gamma$ followed by $\gamma \to e^-+ h^+$, where the lattice $N$ provides the necessary recoil for the creation of an electron-hole pair.   Therefore, we use the vertical transition rate with $(\omega, \mathbf{q}) = (2v_F\Lambda, 0)$ to measure the charge. At the cutoff $\Lambda$, deep inside the band structure and far from the Dirac point, we assume that the electrons behave as in an ordinary insulator and that the effective charge is $e_0^2 \equiv e^2(\widetilde{\mu}) = 4\pi \alpha_\text{EM}/\kappa$.}
The spin degeneracy in Dirac materials is $g_s = 2$; taking $g_C = 1$ (hence $g = 2$) corresponds to a single massive Dirac fermion in QED. The complex permittivity of isotropic semimetals can be recovered from Eq.~(\ref{eq:epsGap}) by taking $\Delta \to 0$ and redefining $\Lambda \to \exp(-5/6)\Lambda$ to absorb the finite $q$-independent piece.  This yields the familiar formula~\cite{AB70,ZCH15,throckmorton2015many,1501.04636,Lv,Rama,2017JPCM...29j5701T}:
\beq\label{eq:epsSM}
\left(\epsilon_r\right)_{\rm semimetal}=1-\frac{e^2g}{24\pi^2 \kappa v_F}\frac{1} {{\bf q}^2}\left\{-\vecq ^2{\rm ln}\left|\frac{4\Lambda^2}{\omega^2/v_F^2-\vecq^2}\right|-i\pi \vecq^2 \Theta(\omega-v_F |{\bf q}|)
\right\}\,,
\eeq
which can also be derived directly from the Lindhard formula, as demonstrated in Appendix~\ref{app:Overlap}.  Eq.~(\ref{eq:epsSM}) was recently confirmed at the 10\% level with optical measurements of Na$_3$Bi~\cite{2016PhRvB..94h5121J}.

Because $\widetilde{\alpha}$ is inversely proportional to $v_F$, materials with small Fermi velocities can have large effective couplings. This is the case of free-standing graphene, where $\kappa = 1$, $v_F = 3 \times 10^{-3}$, and $\widetilde{\alpha} \simeq 2.2$, yet perturbation theory still delivers the right predictions when compared to experiment~\cite{RevModPhys.84.1067}. Since QED flows to a free theory in the IR, perturbation theory remains valid near the Dirac point and far from the cutoff $\Lambda$, so long as no strong coupling phase transitions are crossed.\footnote{In gapless Dirac semimetals, $v_F$ is also renormalized \cite{throckmorton2015many,PhysRevB.86.165127,PhysRevB.87.205138}. We do not consider this subtlety for our benchmark gapless Dirac materials, since in any realistic experiment, the material will be gapped and this issue does not arise.} This is believed to be the case for Dirac materials, which are predicted to be weakly interacting \cite{2017arXiv170804080X}.   Because this is consistent with the current experimental and theoretical consensus in the field~\cite{throckmorton2015many,PhysRevB.86.165127,PhysRevB.87.205138}, we conservatively choose benchmark parameters with $\widetilde{\alpha} < 1$ and assume the validity of perturbation theory at one-loop.

\begin{figure*}[t!]
\begin{center}
\includegraphics[width=0.45\textwidth]{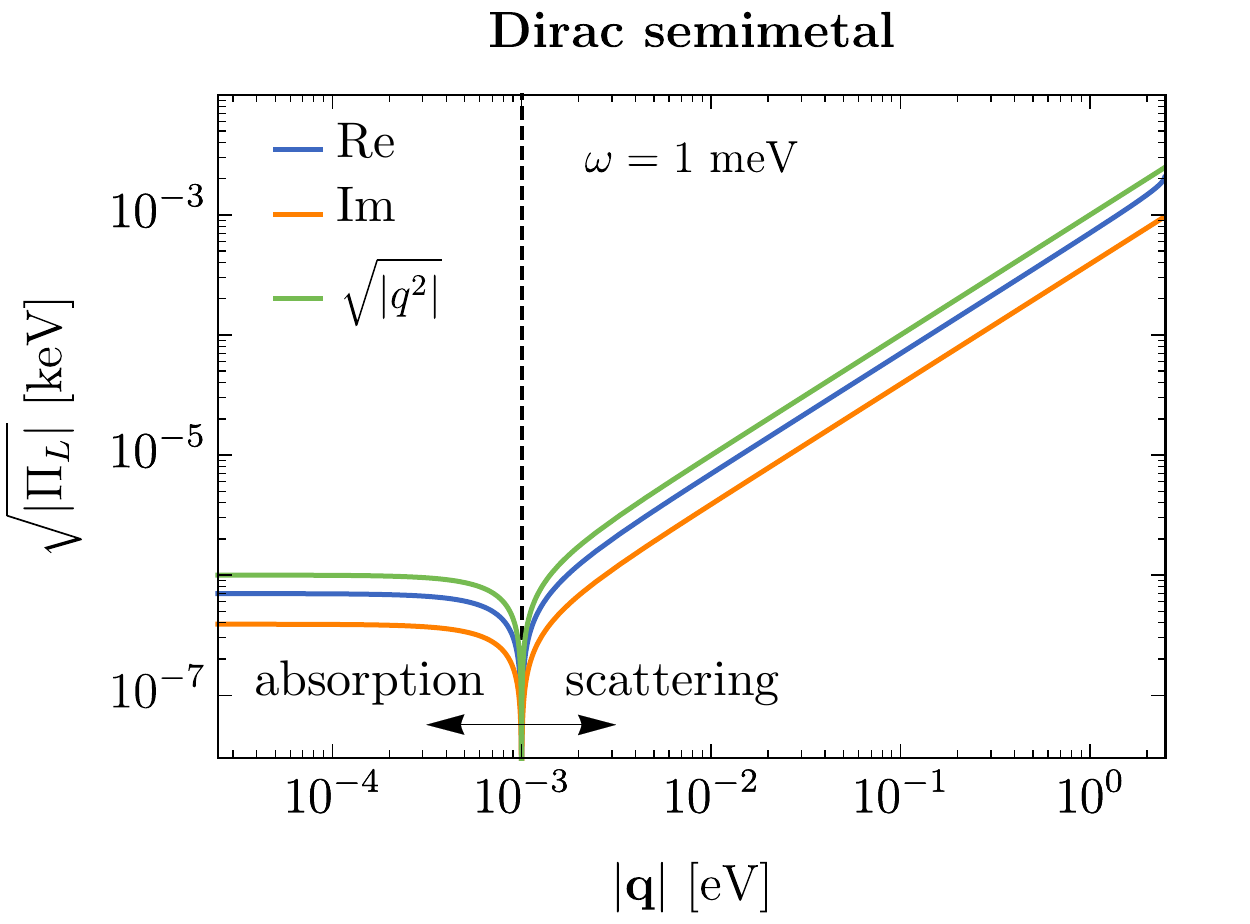}
\includegraphics[width=0.45\textwidth]{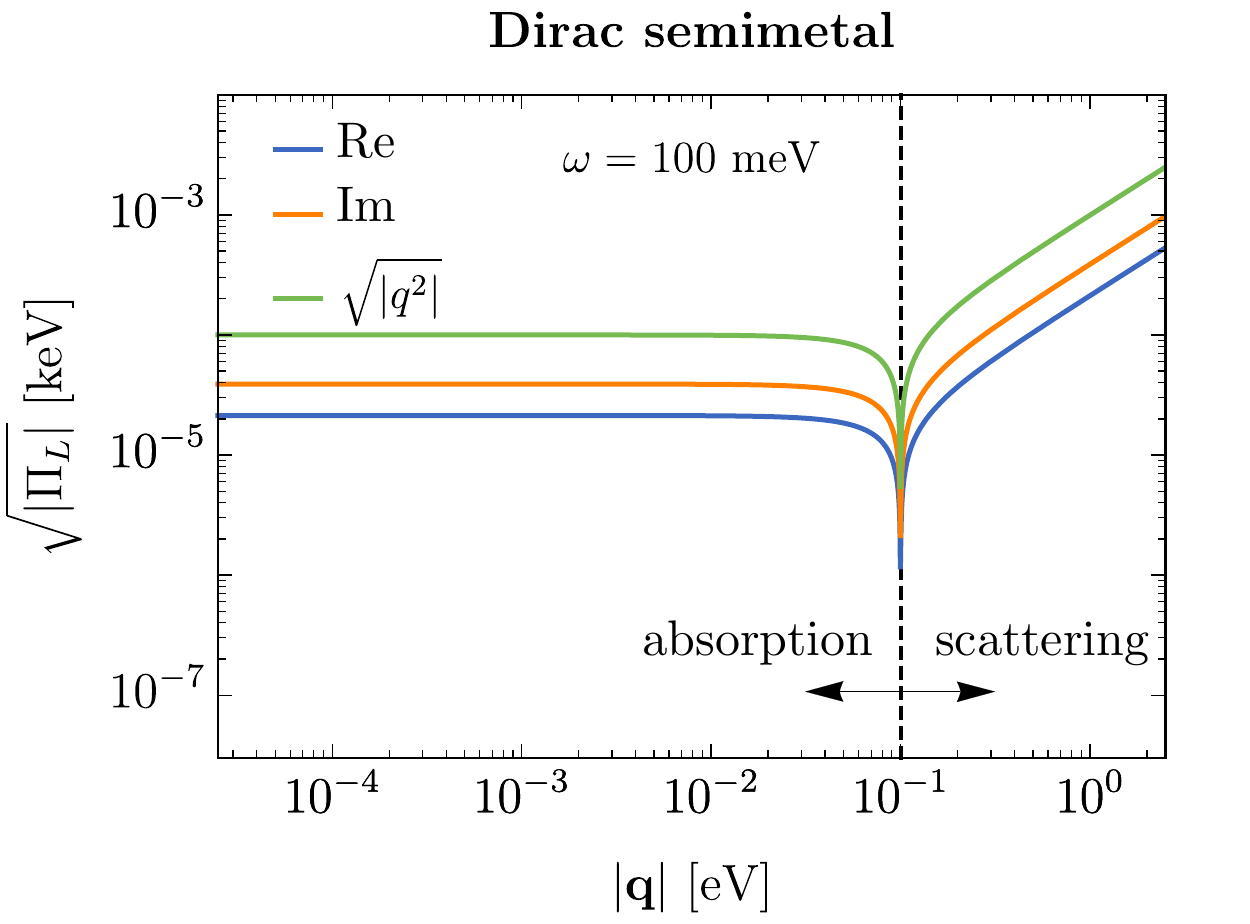}\\
\vspace{.5cm}
\hspace{2.5mm}
\includegraphics[width=0.44\textwidth]{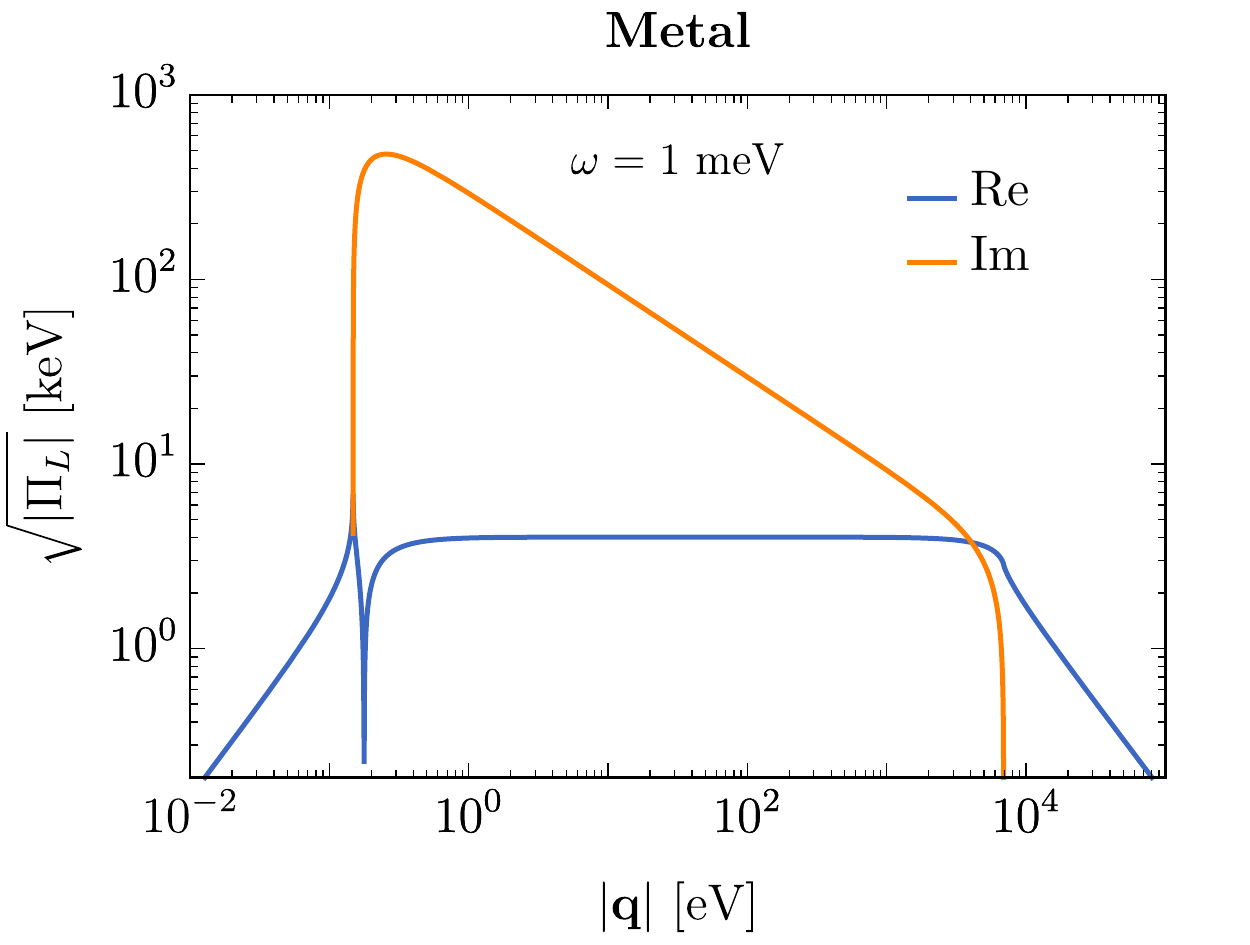}
\hspace{1mm}
\includegraphics[width=0.44\textwidth]{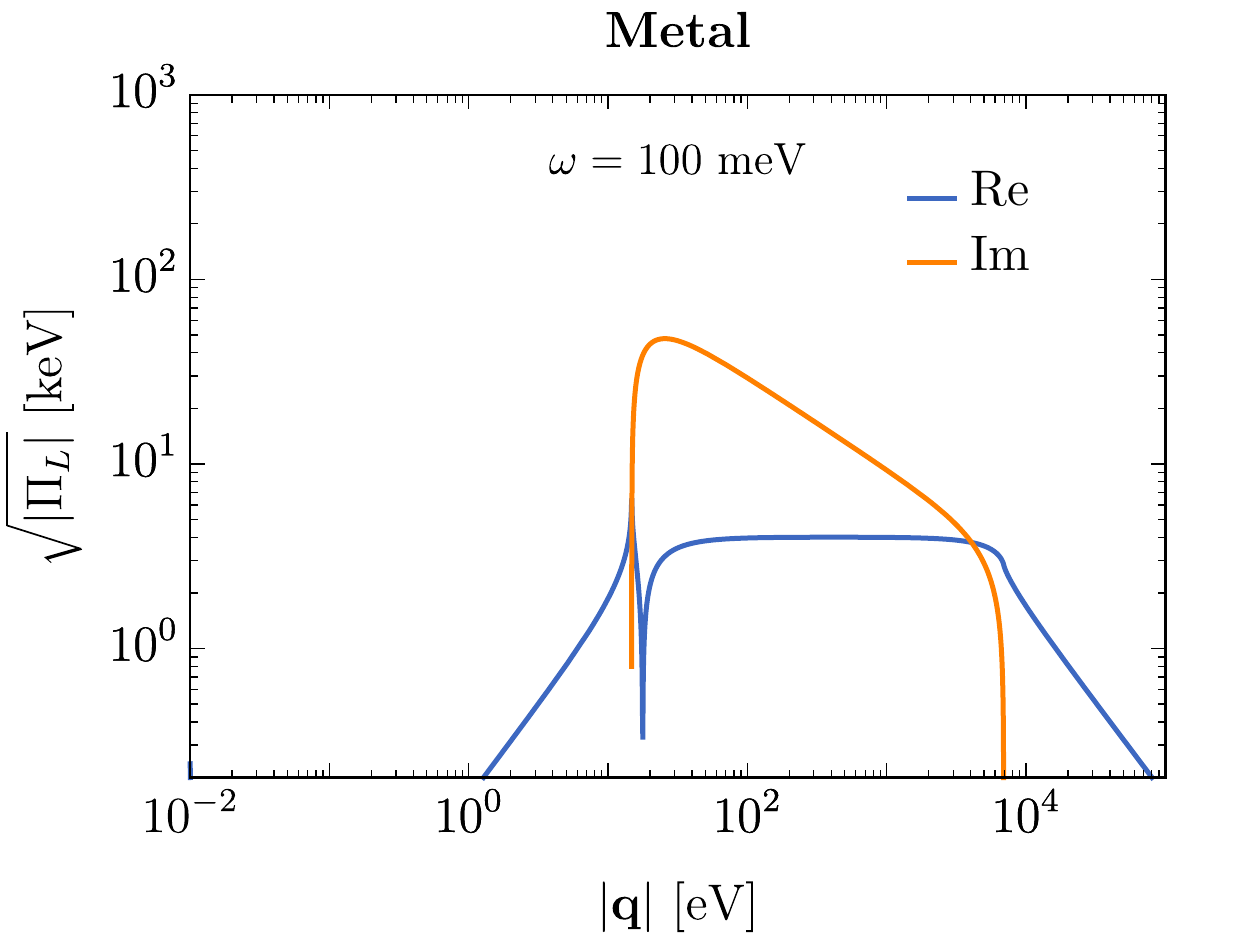}
\caption{ \label{fig:Pi}
The (square root of the) real and imaginary parts of the longitudinal in-medium polarization tensor $\sqrt{|\Pi_L|}$ in Dirac semimetals (\textit{top}) and metals (\textit{bottom}), as a function of the momentum transfer $|\vecq|$.  The left~(right) panel takes the deposited energy to be $\omega=1$~(100)~meV.  For semimetals, we take representative parameters $v_F = 4\times 10^{-4}$, $\Lambda=0.2 \ \keV \simeq 0.1~\ang^{-1}$, $g = 2$, and $\kappa = 40$ to give $\widetilde{\alpha} \sim 0.9$. Note that for semimetals, inter-band transitions are only allowed for $|\vecq| < \omega/v_F$. For metals, we choose aluminum as a representative example, with $\lambda_{\rm TF} \simeq 4 \ \keV$ and $p_F \simeq 3.5 \ \keV$. 
}
\end{center}
\end{figure*}

The permittivity of a Dirac semimetal exhibits distinctive behavior as a function of $\mathbf{q}^2$.  As can be seen from Eq.~\eqref{eq:epsSM}, the imaginary part of $\epsilon_r$ approaches a constant at one-loop order, which is a signature of Dirac-like excitations with linear dispersions~\cite{ZCH15,1501.04636,Lv,Rama,2017JPCM...29j5701T,2016PhRvB..94h5121J}.  The dependence on $\vecq^2$ of the real part of $\epsilon_r$ is mild due to the log, and thus it is also approximately constant.  The top panel of Fig.~\ref{fig:Pi} shows the square-root of the real and imaginary parts of $\Pi_L = q^2 (1-\epsilon_r)$ as a function of $|{\bf q}|$ for $\omega = 1$~(100)~meV in the left~(right) panel.  As a benchmark, we take $v_F = 4 \times 10^{-4}$, $\Lambda = 0.2$~keV, $\kappa = 40$, and $g = 2$, which are representative of typical values for real Dirac materials. The vertical dashed line corresponds to $|{\bf q}| = \omega$; below this point, absorption processes dominate, while scattering processes dominate above it.  To guide the eye, we plot $\sqrt{|q^2|}$ as the solid green line, which scales linearly with $|{\bf q}|$ in the scattering regime $|\vecq| \gg \omega$ and is constant in the absorption regime.  Importantly, the square-root of both the real and imaginary values of $\Pi_L$ track $\sqrt{|q^2|}$, as expected from the fact that $\epsilon_r$ is essentially constant in $\mathbf{q}^2$ for Dirac semimetals.

We discuss modifications to the complex permittivity for anisotropic Dirac materials (where there are independent Fermi velocities, $v_{F,x},~v_{F,y},~v_{F,z}$) in Appendix~\ref{app:aniso}.

\subsection{Comparison of Metals and Dirac Semimetals}
\label{ssec:metalcomparison}

We will show in Sections~\ref{sec:scattering} and~\ref{sec:absorption} that Dirac materials are more sensitive than superconductors to DM scattering via a dark photon mediator, as well as to absorption of dark photons~\cite{Hochberg:2015fth,Hochberg:2016sqx}.  There are competing effects that drive this result.  On the one hand, the optical response of a metal is much stronger than that of a Dirac semimetal, weakening its sensitivity to dark photon interactions.  On the other hand, a metal has a much larger phase space of conduction electrons at low energies, which should improve its reach.  We now discuss the balance of these effects, comparing the optical response and phase space availability in metals versus Dirac semimetals.

For metals, intra-band transitions dominate because the Fermi energy lies within a single band and excitations occur just above the Fermi surface.  In this case, the permittivity is given by:
\begin{equation}\label{ElectricPermittivity}
\left(\epsilon_r\right)_{\rm metal} = 1 +
\frac{\lambda_{\rm TF}^2}{|{\bf q}|^2}\left\{\frac{1}{2} + \left ( \frac{p_F}{4 |{\bf q}|}\left[1 -
\left(\frac{|{\bf q}|}{2 p_F} - \frac{\omega}{|{\bf q}| v_F}\right)^2\right]\ln
\left[\frac{\frac{|{\bf q}|}{2 p_F}-\frac{\omega}{|{\bf q}| v_F} + 1}{\frac{|{\bf q}|}{2
p_F}-\frac{\omega}{|{\bf q}| v_F} - 1}\right] + (\omega \to -\omega)\right) \right\} \, , \end{equation}
where $\lambda_{\rm TF}^2 = 3 e^2 n_e/(2E_F)$ is the Thomas-Fermi screening
length, $n_e$ is the electron density, $p_F$ is the Fermi momentum, and $E_F$ is the Fermi energy~\cite{dressel}.  We plot the square-root of the real and imaginary parts of $\Pi_L$ in a metal in the bottom panel of Fig.~\ref{fig:Pi}. By comparing the top and bottom panels, it is evident that the magnitude of both the real and imaginary components of the polarization tensor are many orders of magnitude smaller in Dirac materials than in metals. Furthermore, the polarization for a metal is roughly constant in $|\mathbf{q}|$ over a broad range of momenta near $\mathcal{O}({\rm keV})$---therefore, we can think of the photon as having an effective mass in this range. By contrast, the real part of the semimetal polarization function scales as $q^2$ up to logarithmic corrections and thus acts as a charge renormalization.

The difference in behavior between the two materials is related to their differing Fermi surface geometries. In metals,
the Fermi surface is not scale invariant; the dimensional Fermi momentum $p_F$ sets the screening scale.
In an (undoped) Dirac semimetal, the Fermi surface is point-like and thus the Fermi momentum is zero by definition. Consequently, there is no screening length for the photon. Alternatively, one can understand this fact from the vanishing of the density of states at the Fermi level in semimetals. The Thomas-Fermi screening length is inversely proportional to the density of states, which is large for a metal and zero for a semimetal. For the case of gapped Dirac materials, one can exploit the emergent Lorentz symmetry of the Dirac Hamiltonian, \Eref{eq:DiracHam}, to see that the Ward identity enforces $\Pi_L(q^2) \sim q^2$ such that the photon stays exactly massless to all orders in perturbation theory; the gap $2\Delta$ does not provide a screening scale akin to $p_F$ in a metal.

As we have just seen, the point-like Fermi surface in a semimetal suppresses its optical response, thereby enhancing processes mediated by a kinetically mixed dark photon.  While this benefits detection rates, it simultaneously suppresses the available phase space for interactions with the DM.  One can use simple geometric arguments to estimate the phase space available for ultra-low-energy scattering in Dirac semimetals compared to metallic targets, for a given energy deposit $\omega$. In a metal, the Fermi surface is a sphere, so the volume of the initial-state phase space is given by a spherical shell of radius $p_F$ and thickness $\delta p = \sqrt{2 m_e \omega}$, where $m_e$ is the electron mass.  Numerically, the phase space volume for $p_F \simeq 3.5 \ \keV$ and $\omega = 1 \ \meV$ is
\be
V_{F, \ \rm{metal}} = 4\pi p_F^2 \,\delta p \sim 5 \times 10^{9} \ \eV^3.
\ee
In a semimetal, the initial-state phase space volume is given by the boundary of the hypercone traced out by the valence band.  The maximum momentum available for the same energy transfer $\omega$ is given by $p_{\rm max} = \omega/v_F$. The phase space volume for $v_F = 4 \times 10^{-4}$ and $\omega = 1 \ \meV$ is
\be
V_{F, \ \rm{Dirac}} = \frac{4}{3}\pi p_{\rm max}^3 \sqrt{1 + v_F^2} \sim  70 \ \eV^3,
\ee
approximately eight orders of magnitude smaller than the corresponding phase space for metals. As shown in Fig.~\ref{fig:Pi}, however, the phase space suppression in the scattering rate is more than offset by the gain from the reduced in-medium response: the scale of the effective dark photon coupling in metals can be 4--6 orders of magnitude larger. When squared, this leads to a huge suppression in the rate, which dominates over the phase space suppression of semimetals. We demonstrate this behavior explicitly in Sections~\ref{sec:scattering} and~\ref{sec:absorption}, where we derive the DM scattering and absorption rates in Dirac materials.

\section{\textbf{Scattering in Dirac Materials}}
\label{sec:scattering}

The formalism for DM scattering in Dirac materials is a special case of the more general formalism for scattering in crystal lattices described in Ref.~\cite{Essig:2015cda}.  We describe the calculation of the DM scattering rate in Section~\ref{ssec:inter} and highlight important issues pertaining to the kinematics in Section~\ref{ssec:ScatteringKinematics}, including the dependence of the scattering rate on the Fermi velocity $v_F$. In Section~\ref{ssec:projections}, we discuss the projected sensitivity to DM scattering in a generic Dirac target and for ZrTe$_5$ in particular.

\subsection{Scattering Rate Formalism}
\label{ssec:inter}
Consider a Dirac cone located at $\vecK$ in the BZ, and a transition from $\veck = \vecK + \vecl$ in the valence band to $\veck' = \vecK + \vecl'$ in the conduction band with $|\vecl|, |\vecl'| \ll |\vecK|$. In order to present simplified analytic results where possible, we assume the gapless, isotropic dispersion relations:
\be
E^{\pm}_\vecl = \pm v_F |\vecl|.
\label{eq:dispersion}
\ee
The main effect of a gap is to impose a kinematic threshold $2\Delta$ on the scattering event, but our conclusions are otherwise unchanged.  A more complete discussion of  anisotropic materials with independent Fermi velocities $v_{F,x},~v_{F,y},~v_{F,z}$ is included in Appendix~\ref{app:aniso}.

The rate to scatter from the valence band (labeled by `$-$') at $\veck$ to the conduction band (labeled by `+') at $\veck'$ is given by \cite{Essig:2015cda}
\be
R_{-,\veck \to +,\veck'} = \frac{\rho_\chi}{m_\chi}\frac{\overline{\sigma}_e}{8\pi \mu_{\chi e}^2}\int d^3 \vecq \, \frac{1}{|\vecq|}\eta\left(v_{\rm min}(|\vecq|,  \omega_{\veck \veck'})\right)|F_{\rm DM}(q)|^2 |\mathcal{F}_{\rm med}(q)|^2 |f_{-,\veck \to +,\veck'}(\mathbf{q})|^2,
\label{eq:Rkkpr}
\ee
where $\rho_\chi \simeq 0.4 \ \GeV/{\rm cm}^3$ is the local DM density, $\mu_{\chi e}$ is the DM-electron reduced mass, $\overline{\sigma}_e$ is a fiducial spin-averaged DM--free-electron scattering cross section, and $\omega_{\veck \veck'}$ is the energy difference between the final and initial states.  The rate also depends on several form factors, which are defined explicitly below: $F_{\rm DM}(q)$ parameterizes the momentum dependence of the DM--free-electron interaction, $\mathcal{F}_{\rm med}(q)$ parameterizes the momentum-dependent in-medium effects, and $f_{-,\veck \to +,\veck'}(\mathbf{q})$ is the transition form factor parameterizing the transition between bands.  Because a distribution of DM velocities contributes to a scattering event with given $\veck, \veck'$, the rate depends on the halo integral:
\be
\eta(\vmin) = \int \frac{d^3 v}{v} \, g_\chi(v) \theta(v - \vmin).
\ee
Here, $g_\chi(v)$ is the DM velocity distribution, which we take to be the Standard Halo Model with typical Galactic-frame velocity $v_0 = 220~{\rm km/s}$ ($7.3 \times 10^{-4}$ in natural units), average Earth velocity with respect to the Galactic frame $v_E = 232~{\rm km/s}$ ($7.8 \times 10^{-4}$), and escape velocity $v_{\rm esc} = 550~{\rm km/s}$ ($1.8 \times 10^{-3}$). For simplicity, we will assume the DM velocity distribution is spherically symmetric. The minimum velocity for a DM particle to scatter with momentum transfer $\vecq$ and energy deposit $\omega_{\veck \veck'}$ is:
\be
v_{\rm min}(|\vecq|,  \omega_{\veck \, \veck'}) = \frac{\omega_{\veck \veck'}}{|\vecq|} + \frac{|\vecq|}{2m_\chi} = \frac{v_F(|\vecl'| + |\vecl|)}{|\vecq|} + \frac{|\vecq|}{2m_\chi} \, .
\label{eq:vmin}
\ee
This expression for $\vmin$ arises from solving a delta function for energy conservation assuming a spherically-symmetric $g_\chi(v)$---see Ref.~\cite{Essig:2015cda} for more details.
Here, we have assumed the gapless isotropic dispersion relation near the $\vecK$-point given in  \Eref{eq:dispersion}; the result generalizes straightforwardly to gapped or anisotropic dispersions.

There are three form factors that appear in Eq.~\eqref{eq:Rkkpr}, two of which are related to the DM scattering interaction and one of which depends on the initial and final wavefunctions of the scattered electron.  We begin by describing the latter.  The transition form factor is defined as
\be
f_{-,\veck \to +,\veck'}(\mathbf{q}) \equiv \int d^3 \mathbf{x} \, \Psi_{+,\veck'}^*(\mathbf{x}) \Psi_{-,\veck}(\mathbf{x}) \, e^{i \mathbf{q}\cdot \mathbf{x}}\,,
\label{eq:fgen}
\ee
where $ \Psi_{-,\veck(+,\veck')}(\mathbf{x})$ is the electron wavefunction in the initial~(final) state. An analytic expression for this factor can be derived using the Hamiltonian in Eq.~(\ref{eq:DiracHam}) and is given by\
\be
|f_{-,\veck \to +,\veck'}(\mathbf{q})|^2 = \frac{1}{2} \frac{(2\pi)^3}{V} \left(1 - \frac{ \vecl \cdot \vecl'}{|\vecl| |\vecl'|}\right) \delta(\vecq - (\vecl' - \vecl)) \,,
\label{eq:TFFAnalytic}
\ee
for gapless isotropic materials, where $V$ is the crystal volume.  A complete derivation of \Eref{eq:TFFAnalytic}, generalized for anisotropic gapped Dirac materials, is provided in Appendix~\ref{app:Overlap}.

The other two form factors, $F_\text{DM}(q)$ and $\mathcal{F}_{\rm med}(q)$, are derived from the matrix element corresponding to a DM particle scattering off an electron via the kinetically mixed dark photon we are interested in:
\begin{equation}
\label{eq:Matrix2}
\langle{|\mathcal {M}|^2}\rangle \simeq \frac{16 m_e^2 m_\chi^2
g_D^2
e^2\varepsilon^2}{\left(q^2-m_{A'}^2\right)^2\left |1-\Pi_L(q)/q^2\right |^2} = \frac{16 m_e^2 m_\chi^2
g_D^2
e^2\varepsilon^2}{\left(q^2-m_{A'}^2\right)^2} \frac{1}{|\epsilon_r(q)|^2} \, ,
\end{equation}
where $g_D$ is the dark photon gauge coupling and $m_e$ is the electron mass. Here, we are neglecting the contribution of $\Pi_T$ to the matrix element, since $\Pi_T \ll \Pi_L$ in the regime $|q^2| \gg \omega^2$ relevant for scattering. The longitudinal polarization tensor $\Pi_L$ (or equivalently, the permittivity $\epsilon_r$) describing the material can thus be incorporated into the event rate for DM scattering using this modified matrix element. We adopt standard conventions in the literature and define $F_{\rm DM}$ as the momentum dependence of the \emph{free} matrix element,
\be\label{eq:FDMdef}
\langle{|\mathcal {M}_{\rm free}(q)|^2}\rangle = \frac{16 m_e^2 m_\chi^2
g_D^2
e^2\varepsilon^2}{\left(q^2-m_{A'}^2\right)^2} \equiv \langle{|\mathcal {M}_{\rm free}(q_0)|^2}\rangle \times |F_{\rm DM}(q)|^2,
\ee
while $\mathcal{F}_{\rm med}$ captures the in-medium effects through
\be
\langle{|\mathcal {M}|^2}\rangle \equiv \langle{|\mathcal {M}_{\rm free}(q)|^2}\rangle \times |\mathcal{F}_{\rm med}(q)|^2.
\ee

The reference momentum $q_0$ used to define $F_{\rm DM}(q)$ in Eq.~\eqref{eq:FDMdef} is arbitrary.  Following the standard of comprehensive reviews such as Ref.~\cite{Battaglieri:2017aum}, we choose $q_0^2 = (\alpha_\text{EM} \, m_e)^2$. Finally, the fiducial cross section is defined as
\be
\overline{\sigma}_e = \frac{\mu_{\chi e}^2}{16\pi m_\chi^2 m_e^2} \langle{|\mathcal {M}_{\rm free}(q_0)|^2}\rangle.
\ee
With these definitions, we have for the light ($m_A' \ll  \ \keV$) kinetically mixed dark photon,
\be
F_{\rm DM}^{A', \, {\rm light}}(q)  = \frac{q_0^2}{q^2}, \qquad \mathcal{F}_{\rm med}(q) = \frac{1}{ \epsilon_r(q)}, \qquad \overline{\sigma}_e = \frac{16\pi \mu_{\chi e}^2 \varepsilon^2 \alpha_{\rm EM} \alpha_D}{q_0^4}\qquad (q_0^2 = (\alpha_{\rm EM} m_e)^2),
\ee
where $\alpha_D = g_D^2/4\pi$ and $\mathcal{F}_{\rm med}(q)$ is evaluated at $q = (\omega_{\vecl \vecl'}, \vecq)$ for initial and final states labeled by $\vecl$ and $\vecl'$ respectively.
Because in Dirac materials $\epsilon_r(q)$ is effectively the ratio of unscreened charge $e_0$ to running charge $e(q)$, the in-medium form factor ensures that the matrix element scales as $e^2(q)$ rather than $e_0^2$.  In Appendix~\ref{app:OtherModels}, we provide the analogous form factor expressions and fiducial cross sections for DM scattering with electrons via other mediators.

The total scattering rate in the crystal is obtained from Eq.~\eqref{eq:Rkkpr} by summing over initial and final states, which in this context means integrating over the initial and final BZ momenta:
\be
R_{\rm crystal} = g_s \, V^2 \int_{\rm BZ} \frac{ d^3 \veck \, d^3 \veck'}{(2\pi)^6} R_{-,\veck \to +,\veck'} = g_s g_C \, V^2 \int_{\rm cone} \frac{ d^3 \vecl \, d^3 \vecl'}{(2\pi)^6} R_{-,\vecl \to +,\vecl'} \, .
\label{eq:Rtot}
\ee
Note that there is no sum over bands because scattering only takes place between the $-$ and $+$ bands by assumption. If there are several Dirac points $\vecK_i$ with identical linear dispersion, one can simply integrate over the region surrounding one of the points and multiply by $g_C$, giving an overall factor of $g = g_s g_C$.  Because the rate only depends on the integral around the cone (we do not consider inter-cone scattering in this paper), and the absolute location of the cone in the BZ is irrelevant, we will work exclusively in terms of the displacement vector $\vecl$ instead of $\veck$ from now on.

\subsection{Scattering Kinematics and Spectrum}
\label{ssec:ScatteringKinematics}
We can exploit the analytic expressions for the transition form factor and the in-medium form factor to analyze the kinematics of scattering in a Dirac material. Using the analytic expression for the transition form factor given in Eq.~(\ref{eq:TFFAnalytic}), we obtain the rate for the case of an ultralight kinetically mixed mediator:
\begin{align}
& R_{-,\vecl \to +,\vecl'} = \frac{\rho_\chi}{m_\chi}\frac{\overline{\sigma}_e}{8\pi \mu_{\chi e}^2} \frac{288 \pi^4 \kappa^2 v_F^2 q_0^4}{e^4 g^2} \frac{(2\pi)^3}{V} \nonumber \\
& \times  \int d^3 \vecq \, \frac{1}{|\vecq|}\frac{1}{(\omega_{\vecl \vecl'}^2 - \vecq^2)^2}\frac{\eta\left(v_{\rm min}(|\vecq|, \omega_{\vecl \vecl'})\right)}{\ln^2 \left|\frac{4\Lambda'^2}{\omega_{\vecl \vecl'}^2/v_F^2-\vecq^2}\right| + \pi^2} \delta(\vecq - (\vecl' - \vecl))\left(1 - \frac{ \vecl \cdot (\vecl + \vecq)}{|\vecl| |\vecl + \vecq|}\right).
\label{eq:RdiffNoGap}
\end{align}
We have defined $\Lambda' = \Lambda \exp(12\pi^2 \kappa v_F/(ge^2))$ to absorb the constant piece in ${\rm Re}\, \epsilon_r$, and dropped the step function in ${\rm Im}\, \epsilon_r$ because all inter-band transitions satisfy $\omega_{\vecl \vecl'} > v_F |\vecq|$. Integrating over $\vecl$ and $\vecl'$ in a region of size $\Lambda$ near the Dirac point as in Eq.~\eref{eq:Rtot}, and noting that the integrand only depends on the magnitudes of $\vecq$ and $\vecl$ and the angle between them, we find the total rate in counts per unit time per unit detector mass:
\be
R_{\rm tot}  = \frac{36 \pi^2}{e^4} q_0^4 \frac{\rho_\chi}{m_\chi} \frac{\overline{\sigma}_e}{\mu_{\chi e}^2} \times n_e V_{\rm uc} \frac{\kappa^2 }{g} v_F^2 I(v_F, \Lambda, m_\chi),
\label{eq:RtotNoGap}
\ee
where $I(v_F, \Lambda, m_\chi)$ has dimensions of momentum---the full expression is provided in Appendix~\ref{app:aniso}.

\Eref{eq:RtotNoGap} is related to \Eref{eq:Rtot} via $R_{\rm tot} = R_{\rm crystal}/M_{\rm crystal}$, where $M_{\rm crystal}$ is the target mass, and $V = N_{\rm uc} V_{\rm uc}$ with $N_{\rm uc}$ the total number of unit cells in the target and $V_{\rm uc}$ the volume of each unit cell.  Then $n_e = N_{\rm uc}/M_{\rm crystal}$ is the number of Dirac valence band electrons per unit mass of target material.  In Eq.~\eref{eq:RtotNoGap}, we have separated the factors that depend on the DM model from those that depend only on the target material.

Of particular interest is the behavior of $I(v_F,\Lambda,m_\chi)$ as a function of $v_F$, as it can suggest the optimal material properties for maximizing detection rates.  Firstly, $I(v_F, \Lambda,m_\chi) = 0$ for large values of $v_F$ due to the peculiar kinematics associated with linear dispersions.  For scattering very close to the Dirac point, the transition form factor \Eref{eq:TFFAnalytic} enforces momentum conservation $\vecq = \vecl' - \vecl$.\footnote{See Appendix~\ref{ssec:latticemom} for a discussion of momentum conservation versus lattice momentum conservation.} Using this relation, \Eref{eq:vmin} becomes
\be
v_{\rm min}(\vecl, \vecl') = v_F\frac{(|\vecl'| + |\vecl|)}{|\vecl' - \vecl|} + \frac{|\vecl' - \vecl|}{2m_\chi}.
\label{eq:vminl}
\ee
The first term is at least $v_F$ by the triangle inequality, and the second term is nonnegative, so we have $v_{\rm min} > v_F$. If $v_F$ is greater than the largest possible DM velocity,
\be
v_{\rm max} = v_E + v_{\rm esc} \simeq 2.6 \times 10^{-3},
\ee
then scattering is \emph{kinematically forbidden} for any (small) $m_\chi$. Therefore, unsuppressed scattering can only occur if the DM is moving \emph{faster} than the electron target.\footnote{We thank Justin Song for pointing out this phenomenon to us.} This is in sharp contrast to the case of superconductors, where the target velocity should \emph{exceed} that of the DM for low energy scattering to occur~\cite{Hochberg:2015pha,Hochberg:2015fth}.

\begin{figure*}[t!]
\begin{center}
\includegraphics[width=0.47\textwidth]{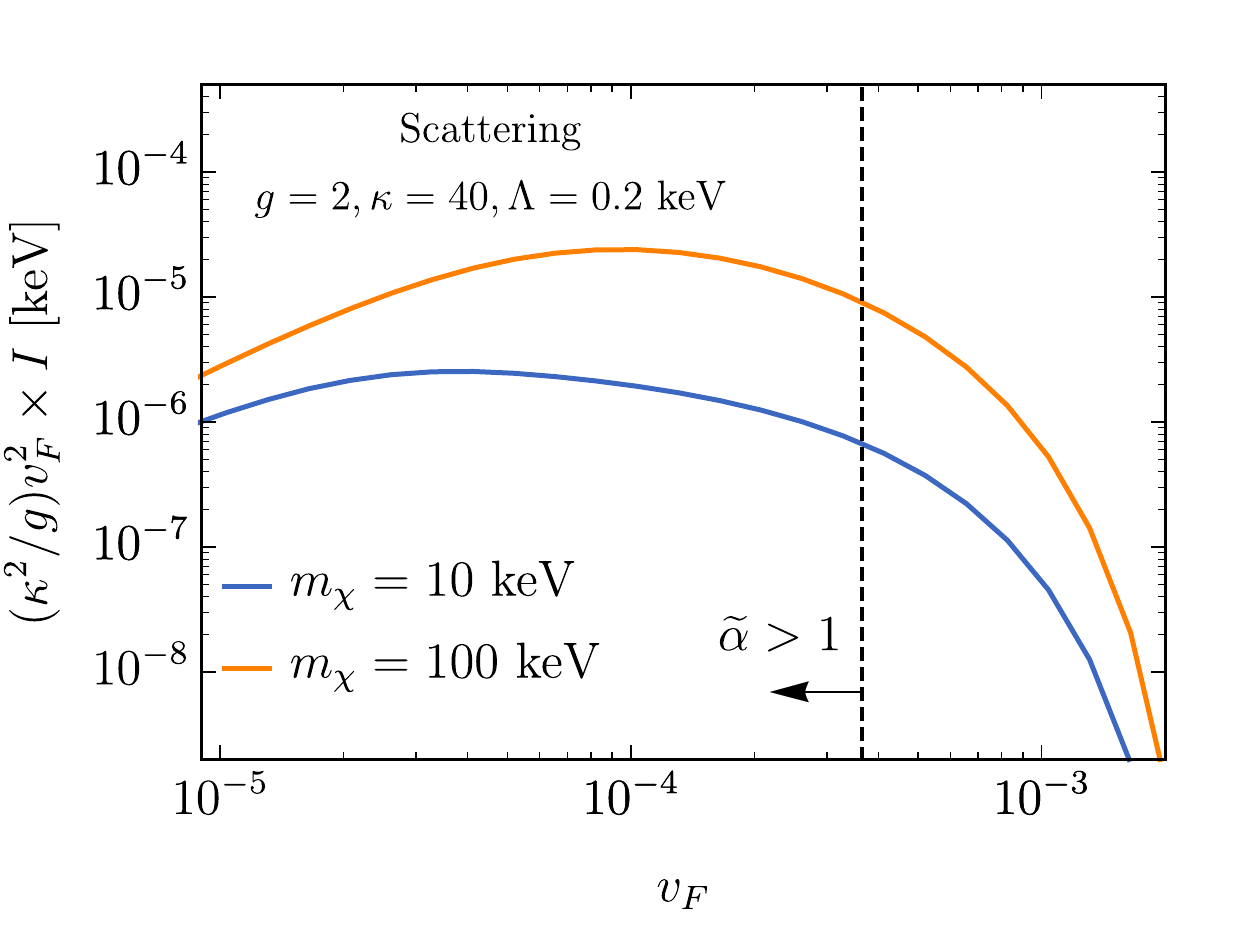}\qquad
\includegraphics[width=0.45\textwidth]{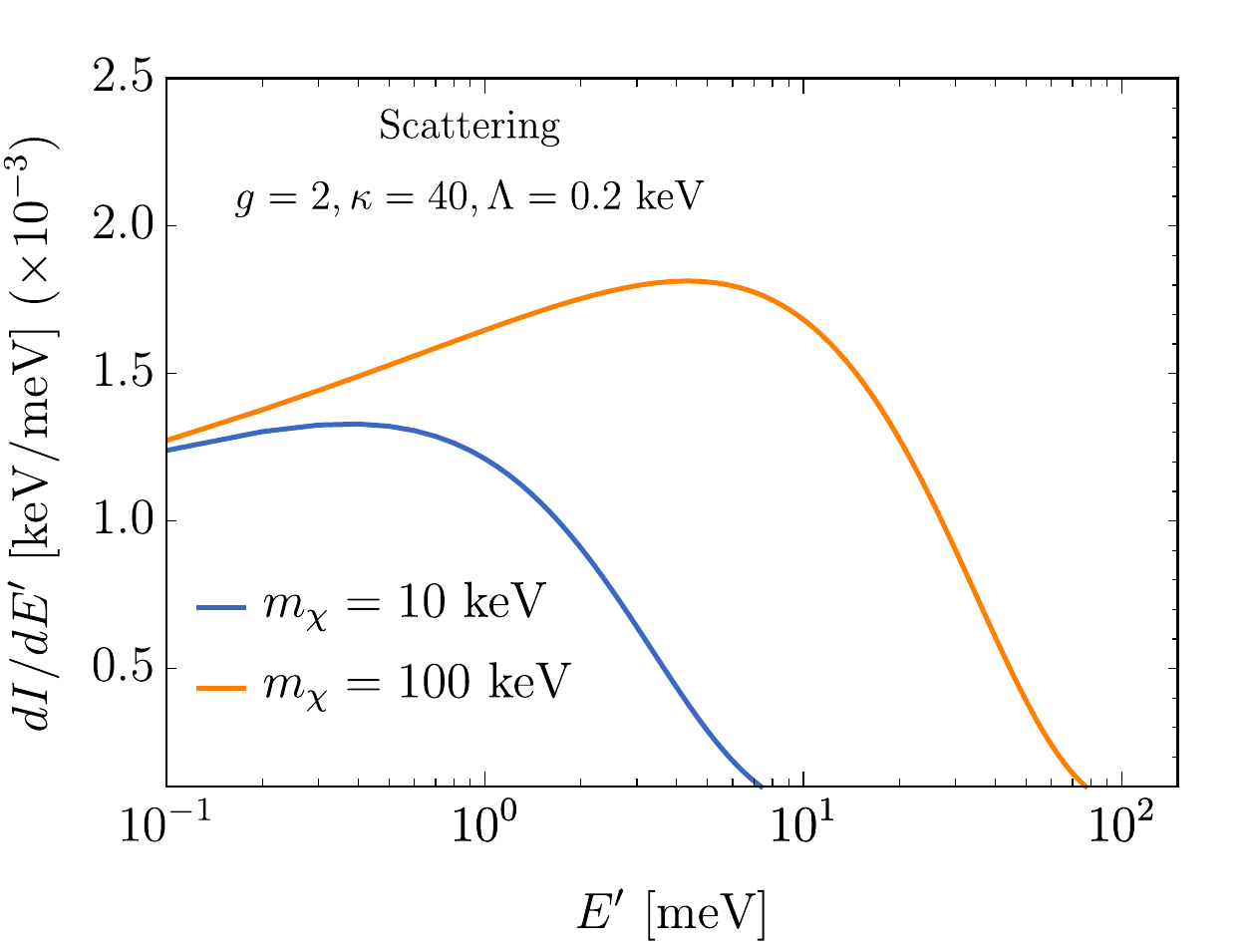}
\caption{ \label{fig:vFSpectrumplot}
(\emph{Left.}) Scaling of the dark matter scattering rate with the Fermi velocity $v_F$ of a gapless isotropic Dirac semimetal.  The vertical dashed line indicates the point below which $\tilde{\alpha}$, the effective fine-structure constant in the medium, is greater than 1.  (\emph{Right.}) Spectrum $dI/dE'$ for $v_F = 4 \times 10^{-4}$, where $E'$ is the final-state energy of the scattered electron. Note that the function $I(v_F, \Lambda, m_\chi)$ is directly proportional to the total scattering rate, $R_\text{tot}$.  In both cases, we have taken $\Lambda = 0.2 \ \keV$, $g = 2$, and $\kappa = 40$.  The results are shown for 10~keV and 100~keV dark matter in blue and orange, respectively.}
\end{center}
\end{figure*}

Dirac materials exhibit a range of Fermi velocities from $6 \times 10^{-3}$ for BLi to $3 \times 10^{-5}$ or smaller for NbAs and NbP \cite{2014arXiv1412.2607D,lee2015fermi}.  The  kinematic arguments presented above suggest that the materials with smallest $v_F$ are most desirable for maximizing the DM scattering rate.  However, the prefactor in Eq.~\eref{eq:RtotNoGap} is suppressed by $v_F^2$, which comes from the scaling of $\epsilon_r$.  Therefore, we do not want to drive $v_F$ too low.  To illustrate this tension, the left panel of Fig.~\ref{fig:vFSpectrumplot} plots $\frac{\kappa^2}{g} v_F^2 I(v_F, \Lambda, m_\chi)$, which is proportional to the total scattering rate, for two values of the DM mass.  The results are shown assuming $\Lambda = 0.2 \ \keV$, $g = 2$, and $\kappa = 40$, representative of typical values for real Dirac materials.  For both masses, the rate is maximal for a particular choice of the Fermi velocity.  When $m_\chi = 100 \ \keV$, this occurs at $v_F \simeq  10^{-4}$.  For $m_\chi = 10 \ \keV$, the maximum is at even lower Fermi velocities.  Such small values for the Fermi velocity lead to $\widetilde{\alpha} > 1$ for the material parameters assumed here.  That said, the rate for either mass point only varies by a factor of a few between the $v_F$ that maximizes the rate and $v_F = 3.6 \times 10^{-4}$, above which the effective coupling is less than 1.

Finally, we consider the energy spectrum $dI/dE'$ of the excited electron, shown in Fig.~\ref{fig:vFSpectrumplot}~(\emph{right}) for $v_F = 4 \times 10^{-4}$ and $m_\chi = 10, 100 \ \keV$. The spectrum peaks away from $E' = 0$ due to the vanishing phase space at the point of the Dirac cone.  This shows that the majority of the rate comes from final-state energies above 1 meV. At small $E'$, the spectrum depends only weakly on $m_\chi$. This is because the energetically favorable events correspond to small initial-state energies, such that $|\vecl' - \vecl|$ is small and $v_{\rm min}$    is approximately independent of $m_\chi$.  As expected, heavier DM masses yield scattering events with higher-energy final-state electrons, giving a larger total rate. As we will show in Section~\ref{ssec:projections} below, these conclusions do not change for $m_{\chi} \gtrsim 10 \ \keV$ even in the presence of a meV-scale gap.

\subsection{Projected Sensitivity Reach}
\label{ssec:projections}

\begin{figure*}[t!]
\begin{center}
\includegraphics[width=0.7\textwidth]{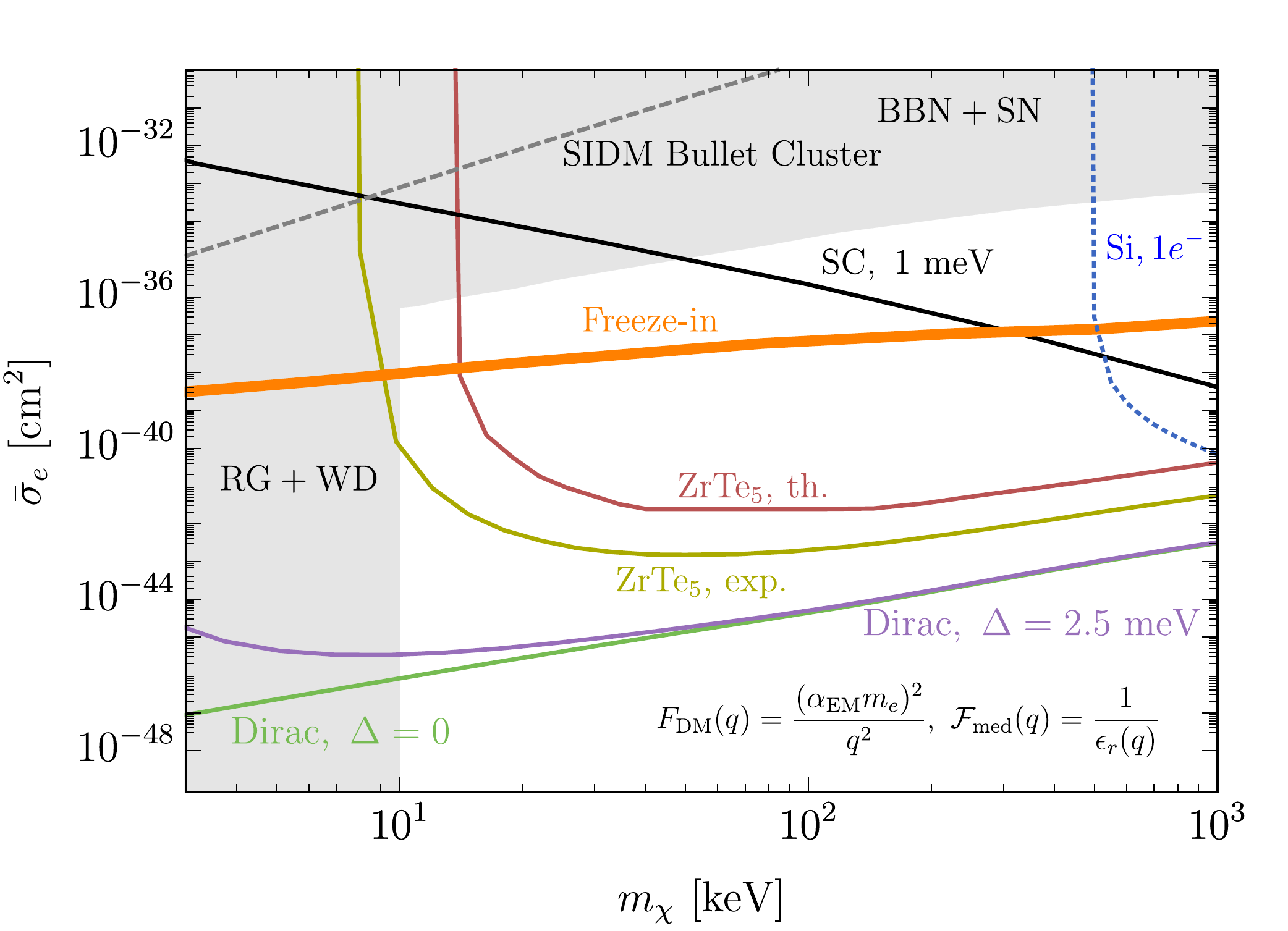}
\caption{ \label{fig:SMkinmix}
Projected reach of dark matter scattering in Dirac materials through a light kinetically mixed dark photon mediator with in-medium effects included.  We show the expected background-free 95\% C.L. sensitivity (3.0 events) that can be obtained with 1 kg-yr exposure. For the two curves labeled `Dirac,' we assume an ideal gapless ($\Delta = 0$, green) or gapped ($\Delta = 2.5 \ \meV$, purple) isotropic Dirac material with $v_F = 4 \times 10^{-4}$, $\kappa = 40$, $g = 2$, $\Lambda = 0.2 \ \keV$, $n_e = 5 \times 10^{24}/{\rm kg}$, and $V_{\rm uc} = 60 \ \ang^3$.  We also the show the results for ZrTe$_5$, a realistic target material.  The red curve labeled `ZrTe$_5$, th.' uses the parameters calculated in Appendix~\ref{app:DFT}, while the yellow curve labeled `ZrTe$_5$, exp.' uses parameters extracted from experiment \cite{Zheng2016,XSY17}. For comparison, we also show the reach of superconductors with a 1 meV threshold~\cite{Hochberg:2015fth} (black), and the projected single-electron reach for a silicon detector with a $1e^{-}$ threshold \cite{Battaglieri:2017aum} (blue dotted). The orange curve labeled `Freeze-in' delineates where freeze-in production~\cite{Hall:2009bx} results in the correct dark matter relic abundance.  The gray shaded regions indicate bounds from white dwarfs, red giants, big bang nucleosynthesis, and supernovae, which are derived from limits on millicharged particles~\cite{Davidson:2000hf,Essig:2015cda}. The gray dashed line indicates bounds on self-interacting dark matter derived from observations of the Bullet Cluster \cite{Feng:2009hw,EssigPrivate}.}
\end{center}
\end{figure*}

We are now ready to use the formalism we developed to present the sensitivity reach projections for DM scattering in Dirac materials via a light kinetically mixed dark photon.  The results are shown in Fig.~\ref{fig:SMkinmix}. The green and purple curves show the expected 95\% C.L. sensitivity (corresponding to 3.0 signal events) with a kg-year exposure for DM scattering in gapless and gapped Dirac materials, respectively.  For concreteness, we choose $\Lambda = 0.2 \ \keV$, $V_{\rm uc} = 60 \ \ang^3$, and $n_e = 5 \times 10^{24}/{\rm kg}$, typical of experimentally realized semimetals. In addition, we take $v_F =4 \times 10^{-4}$, $\kappa = 40$ and $g = 2$ so that $\widetilde{\alpha}  \sim 0.9$ and perturbation theory is reliable.  This corresponds to a typical range of parameters for Dirac semimetals such as Cd$_3$As$_2$ \cite{neupane2013observation,LJZ14,borisenko2014experimental,zivitz1974optical,jay1977electron}, for which perturbation theory is only expected to break down at $\widetilde{\alpha} \simeq 9.4$ \cite{throckmorton2015many}.  
We note that the inclusion of the correct wavefunction overlaps from \Eref{eq:TFFAnalytic} suppresses the rate by about an order of magnitude compared to a naive approximation where the transition form factor is set to unity.

In Fig.~\ref{fig:SMkinmix}, we also show the projected sensitivity for a benchmark realistic target material, ZrTe$_5$, which has most of the desired properties we have discussed. The band structure was determined using density functional theory, as discussed in Appendix~\ref{app:DFT}.  We find that ZrTe$_5$ has a small Fermi velocity $v_{F,y} = 4.9 \times 10^{-4}$ along one direction, a small degeneracy $g = 4$, and a small gap $2\Delta = 35 \ \meV$ at zero temperature. The remaining material parameters are given in Appendix~\ref{app:DFT}. The band structure of ZrTe$_5$ is highly anisotropic, with $v_{F,y} \ll v_{F,x},v_{F,z}$. The crystal lattice also has a highly anisotropic background dielectric tensor, with $\kappa_{yy} \ll \kappa_{xx}, \kappa_{zz}$; we take the harmonic mean $\widetilde{\kappa} = \frac{3}{1/\kappa_{xx} + 1/\kappa_{yy} + 1/\kappa_{zz}} = 25.3$ for our estimates here, and justify this approximation in the context of our assumption of a spherically-symmetric DM distribution in Appendix~\ref{ssec:anperm}. Note that the combined effect of these anisotropies may result in interesting directional dependence of the signal, including daily modulations of the rate, but this requires a dedicated analysis which is beyond the scope of this paper. The effective fine-structure constant is $\widetilde{\alpha} = g \alpha_{\rm EM}/\widetilde{\kappa}(v_{F,x} v_{F,y} v_{Fz})^{1/3} \simeq 0.80$. High-purity ZrTe$_5$ can be synthesized in macroscopic quantities, and pressure or doping can shift the Fermi level inside the gap so that the conduction band is empty at zero temperature. As shown in Refs.~\cite{Hochberg:2015pha,Hochberg:2015fth}, an meV-scale gap with little or no occupation of excited states is necessary for suppressing thermal noise. Experimental measurements of the properties of ZrTe$_5$ have led to some ambiguous results regarding the precise values of the Fermi velocities and $\Delta$, so for comparison, we also plot the projected sensitivity using the measurements of Fermi velocities from Ref.~\cite{Zheng2016}, and a gap energy $2\Delta = 23.5 \ \meV$, the median of the range of values found in Ref.~\cite{XSY17}.

For comparison, we provide the projections for a superconducting target with a 1~meV threshold~\cite{Hochberg:2015fth} (solid black line) and semiconductor target~\cite{Battaglieri:2017aum} (blue dotted line).  For the latter, we show a silicon target with a single-electron threshold. Both are low-threshold electron-scattering experimental proposals with complementary detection modalities: the superconductor proposal exploits the breaking of Cooper pairs to produce quasiparticles and athermal phonons from meV energy deposits, and the semiconductor proposal aims to detect valence-to-conduction excitation (as we propose here) in a generic band structure with a large gap of 1.11 eV. As we have discussed, the reach of Dirac materials is superior to that of superconductors for the case of a light kinetically mixed dark photon mediator due to the reduced in-medium effects. Assuming the DM velocity distribution is given by the Standard Halo Model, semiconductors are unable to probe DM lighter than 500~keV due to their large band gaps.

The orange line in Fig.~\ref{fig:SMkinmix} shows the theory expectation for a benchmark model where the DM abundance is set through freeze-in via a light mediator~\cite{Hall:2009bx}.  In such models, the DM is very weakly coupled to the Standard Model such that it never thermalizes, and the DM abundance is instead gradually populated through very rare interactions at low temperatures.  If these interactions are with the electron, as is the case for DM coupling to a dark photon, freeze-in production gives a concrete theoretical target for electron scattering direct detection experiments.

The constraints on light dark photons can be quite stringent; the excluded regions of parameter space (at least for the most naive of models) are indicated by the gray regions in Fig.~\ref{fig:SMkinmix}. These are derived from bounds on millicharged particles~\cite{Davidson:2000hf}, which are also applicable to DM coupled to an ultralight kinetically mixed dark photon~\cite{Essig:2015cda,Hochberg:2015fth}. When the DM is lighter than the temperature of red giants and white dwarfs, DM can be copiously produced and lead to excessive cooling; in Fig.~\ref{fig:SMkinmix}, this (approximate) region is shown in gray and marked `RG+WD'.  In addition, the presence of dark photons affects the energetics of supernovae and big bang nucleosynthesis (BBN), implying that $\varepsilon^2 \alpha_D \lesssim 10^{-17}$; in Fig.~\ref{fig:SMkinmix}, this region is shown in gray and is marked `SN+BBN'.   Constraints from DM self-interactions are generally weaker; for example, the self-interacting DM bound from observations of the Bullet Cluster~\cite{Feng:2009hw,EssigPrivate}
(labeled `SIDM Bullet Cluster') are subdominant to the other constraints.

The light kinetically mixed mediator scenario we have considered here is particularly interesting for direct detection with Dirac materials because the scattering rate is greatly enhanced at low momentum transfer due to the $1/q^4$ dependence of the DM form factor $|F_{\rm DM}|^2$. Since this momentum dependence is not spoiled by the in-medium form factor $\mathcal{F}_{\rm med}$, Dirac materials are able to probe very small couplings, which are unconstrained by any other observations, cosmological or otherwise. As anticipated in Section~\ref{sec:formalism}, Dirac materials have superior reach in this case to both superconductors, which suffer from an in-medium suppression at low masses, and semiconductors, which have eV-scale gaps. Ideal Dirac materials with small Fermi velocity $v_F \sim 4 \times 10^{-4}$ and small gap $2\Delta = 5 \ \meV$, with 1 kg-yr of exposure, can probe cross sections many orders of magnitude smaller than the entire freeze-in region below 1 MeV. Realistic materials such as ZrTe$_5$ still give excellent reach, which can be improved by identifying materials with smaller Fermi velocities and gaps.

In Appendix~\ref{app:OtherModels}, we present the reach of Dirac materials for DM scattering via a heavy kinetically mixed dark photon, as well as via a light or heavy scalar mediator where no in-medium effects arise. For the former case, we find that Dirac materials provide better sensitivity than superconductors; in the latter case, Dirac materials generally fare worse than superconductors, as expected. Strong constraints from either stellar emission (light mediators) or BBN (heavy mediators) apply at least for the most naive of such models, such that typically either BBN or stellar emission bounds must be evaded for the models where DM does {\em not} scatter via a light dark photon. Our results here demonstrate, however, that Dirac materials are an ideal target for light dark photon mediators.

\section{\textbf{Absorption in Dirac Materials}}
\label{sec:absorption}

Having demonstrated that Dirac materials have compelling reach for the case of DM scattering, we move on to the case of DM absorption.  We begin by presenting the formalism for calculating DM absorption rates, and then discuss the relevant kinematics and projected sensitivities for Dirac materials.

\subsection{Absorption Rate Formalism}
The rate for DM absorption in counts per unit time per unit detector mass is given by
\beq
R_{\rm abs}=\frac{1}{\rho_T} \frac{\rho_{\rm \chi}}{m_{\rm \chi}}\langle n_T \sigma_{\rm abs} v_{\rm rel}\rangle_{\rm DM}\,,
\eeq
where $\rho_T$ is the mass density of the target, $n_T$ is the number of target particles, $\sigma_{\rm abs}$ is the DM absorption cross section on the target, and $v_{\rm rel}$ the relative velocity between the DM and the target. One can relate the absorption rate of certain classes of DM particles to the measured optical properties of the target~\cite{Bloch:2016sjj,Hochberg:2016ajh,Hochberg:2016sqx}. In particular, the absorption rate of photons in a given (bulk) material is determined by the polarization tensor via the optical theorem:
\beq\label{eq:abs}
\langle n_T \sigma_{\rm abs} v_{\rm rel}\rangle_\gamma = -\frac{{\rm Im}\,\Pi(\omega)}{\omega}\,,
\eeq
where $\omega$ is the energy of the incoming absorbed photon and $\Pi(\omega)$ denotes the polarization tensor, in an isotropic material, in the relevant limit of $|{\bf q}|\ll \omega$.   For absorption of DM particles, the deposited energy $\omega$ in the system is equal to the DM mass $m_\chi$, and the momentum transfer $\vecq$ is equal to the DM momentum $m_\chi {\bf v}_{\rm DM}$. Consequently, the momentum transfer is suppressed due to the virial velocity of the DM, $|{\bf q}|\sim 10^{-3}\omega\ll \omega$.  In this limit, $\Pi_L\approx \Pi_T \equiv \Pi$.  Using Eq.~\eqref{PiLPiT}, we can write the absorption rate for photons as
\beq\label{eq:abssigma}
\langle n_T \sigma_{\rm abs} v_{\rm rel}\rangle_\gamma = \omega\; {\rm Im}\,\epsilon_r \,.
\eeq
The sensitivity of a material to DM absorption is therefore obtained by relating the absorption process to that of ordinary photons through the complex permittivity.

We focus on the case of a kinetically mixed dark photon, as described by \Eref{eq:kinmixing}, which can be absorbed by a Dirac material.  Such a dark photon can comprise all of the DM, with its relic abundance set via a misalignment mechanism~\cite{Nelson:2011sf,Arias:2012az,Graham:2015rva}. The effective mixing angle between the dark photon and the photon for the case of absorbtion of non-relativistic DM in the target is given by
\beq\label{eq:kappaeff}
\varepsilon^2_{\rm eff} =  \frac{ \varepsilon^2 m_{A'}^4}{ \left[ m_{A'}^2 - {\rm Re}\;\Pi(m_{A'}) \right]^2 + \left[ {\rm Im}\;\Pi(m_{A'}) \right]^2} \,,
\eeq
and so the rate of absorption is
\beq
	R_{\rm abs}^{A'}=  \frac{1}{\rho_T}  \rho_\chi \varepsilon_{\rm eff}^2 \,  {\rm Im}\,\epsilon_r \,.
\eeq
For dark photon DM in the mass range of meV to hundreds of meV, the energy deposited in absorption matches the regime of interest for Dirac materials.

\subsection{Absorption Kinematics and Scaling}

\begin{figure*}[th!]
\begin{center}
\includegraphics[width=0.48\textwidth]{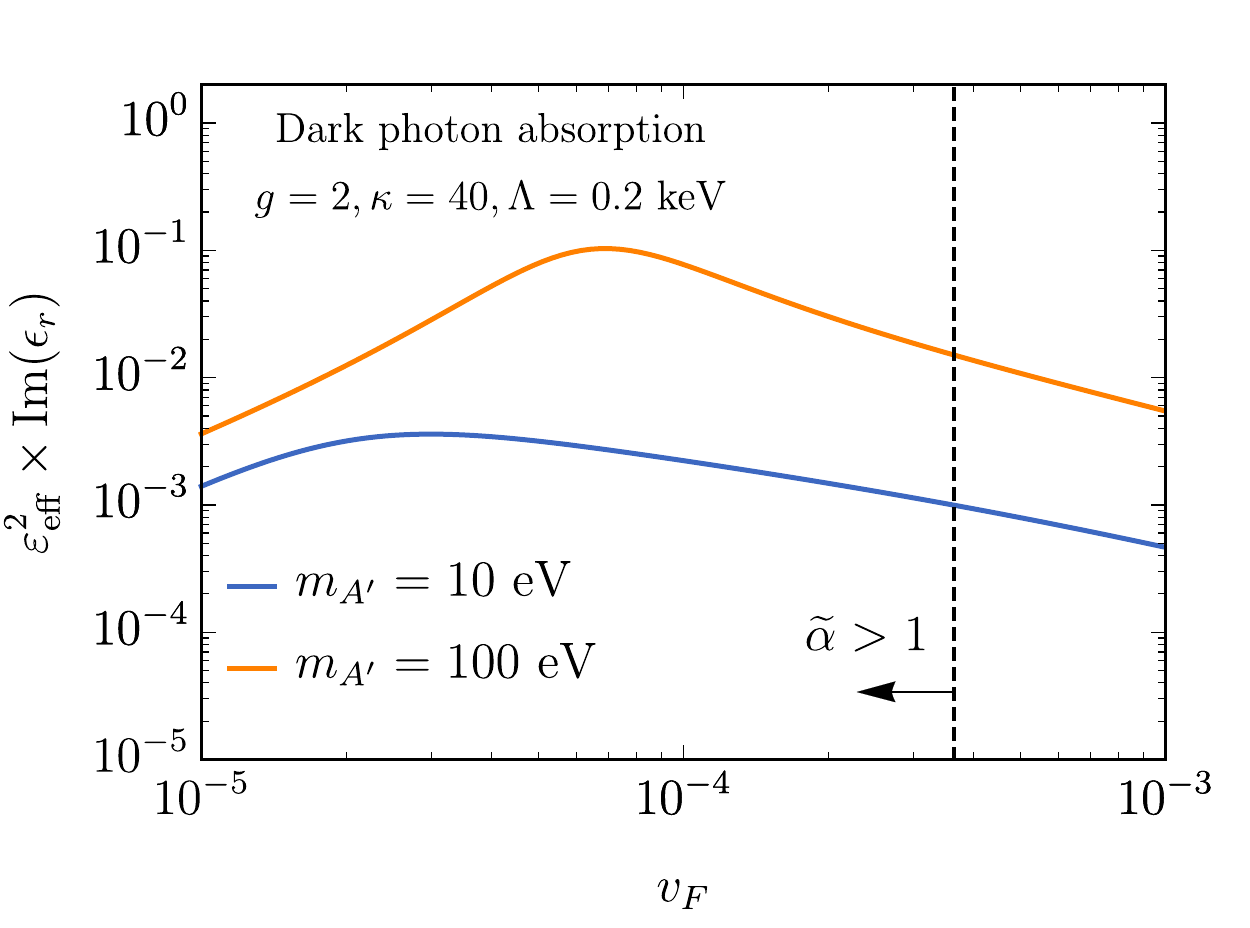}
\includegraphics[width=0.48\textwidth]{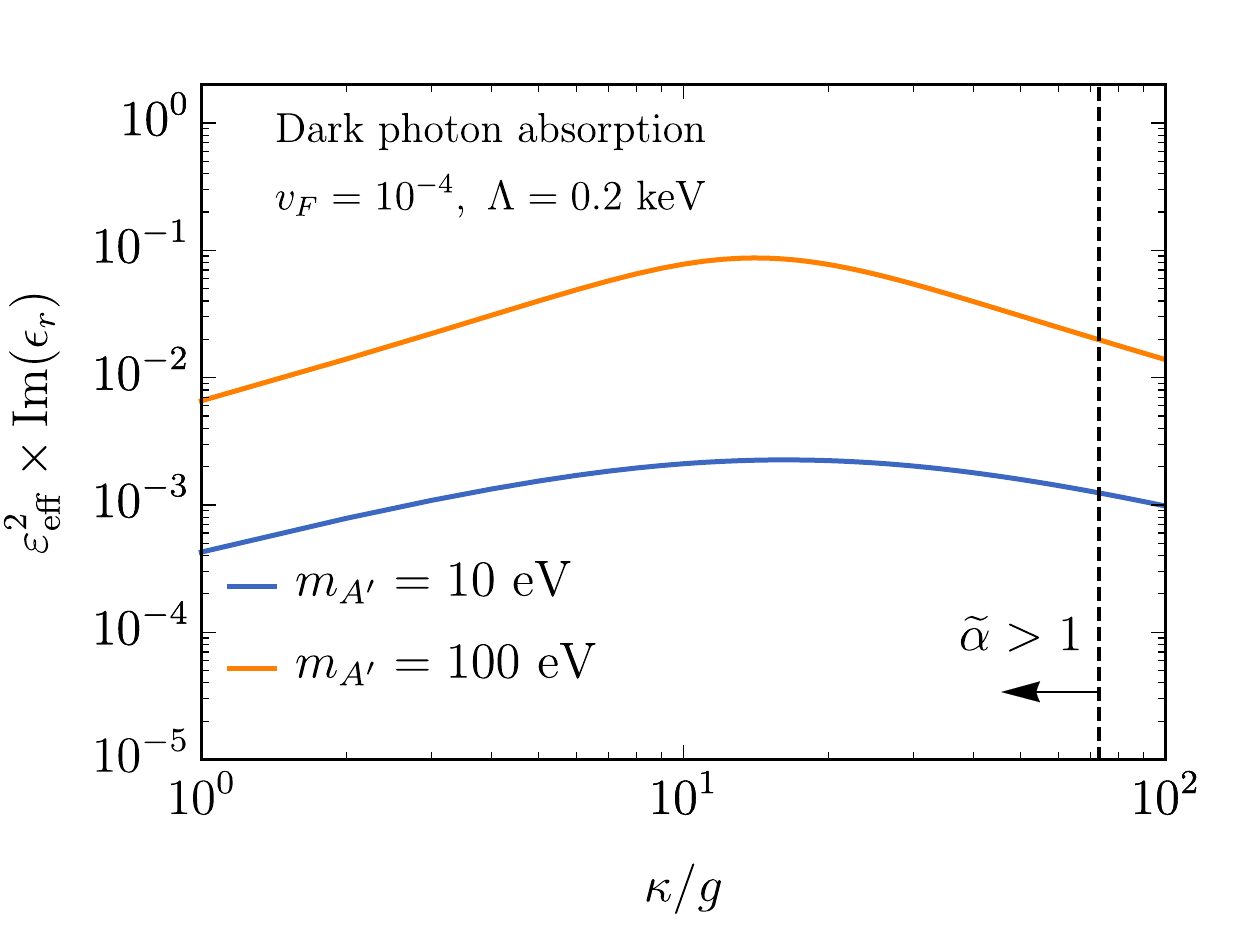}
\caption{ \label{fig:vFabs}
(\textit{Left.}) Scaling of the dark photon absorption rate with $v_F$ for fixed $\kappa/g$. (\textit{Right.}) Scaling of the dark photon absorption rate with $\kappa/g$ for fixed $v_F$. Note that the absorption rate is proportional to $\varepsilon_{\rm eff}^2 \times {\rm Im}(\epsilon_r)$. In both cases we have considered a gapless isotropic Dirac semimetal with $\Lambda = 0.2 \ \keV$ for two values of the dark photon mass, 10 eV (blue) and 100 eV (orange). The dashed line marks where the effective coupling becomes strong, $\widetilde{\alpha} = 1$.
 }
\end{center}
\end{figure*}

As shown in Fig.~\ref{fig:Pi}, $\Pi(q)$ has a non-vanishing imaginary part even at $\vecq = 0$  in a Dirac material. Indeed, this can be considered a distinctive property of Dirac materials~\cite{ZCH15,1501.04636,Lv,Rama,2017JPCM...29j5701T,2016PhRvB..94h5121J}. The physical interpretation is that a Dirac material can absorb an incoming particle with momentum transfer much smaller than its mass, without the presence of additional particles (such as phonons). In other words, vertical transitions from the valence band to the conduction band are possible.  This is in contrast to absorption in typical metals, where inter-band transitions can be neglected for ultralow energies and non-relativistic absorption proceeds through emission of a phonon~\cite{Hochberg:2016ajh}---a process which is not described by the polarization tensor of Eq.~\eqref{ElectricPermittivity}.

Furthermore, because ${\rm Im}\ \epsilon_r$ scales as $\widetilde{\alpha} = \alpha_\text{EM} g/\kappa v_F$, one might expect the absorption rate to increase with small Fermi velocity $v_F$, enhanced degeneracy $g$, and small background dielectric constant $\kappa$. This is indeed the case for absorption of a light scalar or pseudoscalar, which does not feel in-medium effects.
In the case of the kinetically mixed dark photon, \Eref{eq:kappaeff} shows that the effective in-medium mixing angle between the dark photon and the photon involves both real and imaginary parts of $\Pi$, leading to a more complicated dependence on the Dirac material parameters. In Fig.~\ref{fig:vFabs} we show the combination $\varepsilon^2_{\rm eff}\times {\rm Im}(\epsilon_r)$ for a Dirac semimetal, which is proportional to the dark photon absorption rate, for two values of the mass, $m_{A'}=10$ and 100~eV. The left panel fixes $\kappa/g$ and varies $v_F$, while the right panel fixes $v_F$ and varies $\kappa/g$. The optimal Fermi velocity value is mass-dependent, and is of order $10^{-4}$ or smaller, while the optimal $\kappa/g$ is $\mathcal{O}(10)$ for $v_F = 10^{-4}$ and relatively insensitive to $m_{A'}$. However, for these optimal parameters, $\widetilde{\alpha} > 1$. To be conservative, in what follows we will present results using the same Dirac material parameters as in Section~\ref{ssec:projections} such that $\widetilde{\alpha} < 1$ and perturbation theory remains valid at one-loop.

\subsection{Projected Sensitivity Reach}

\begin{figure*}[t!]
\begin{center}
\includegraphics[width=0.65\textwidth]{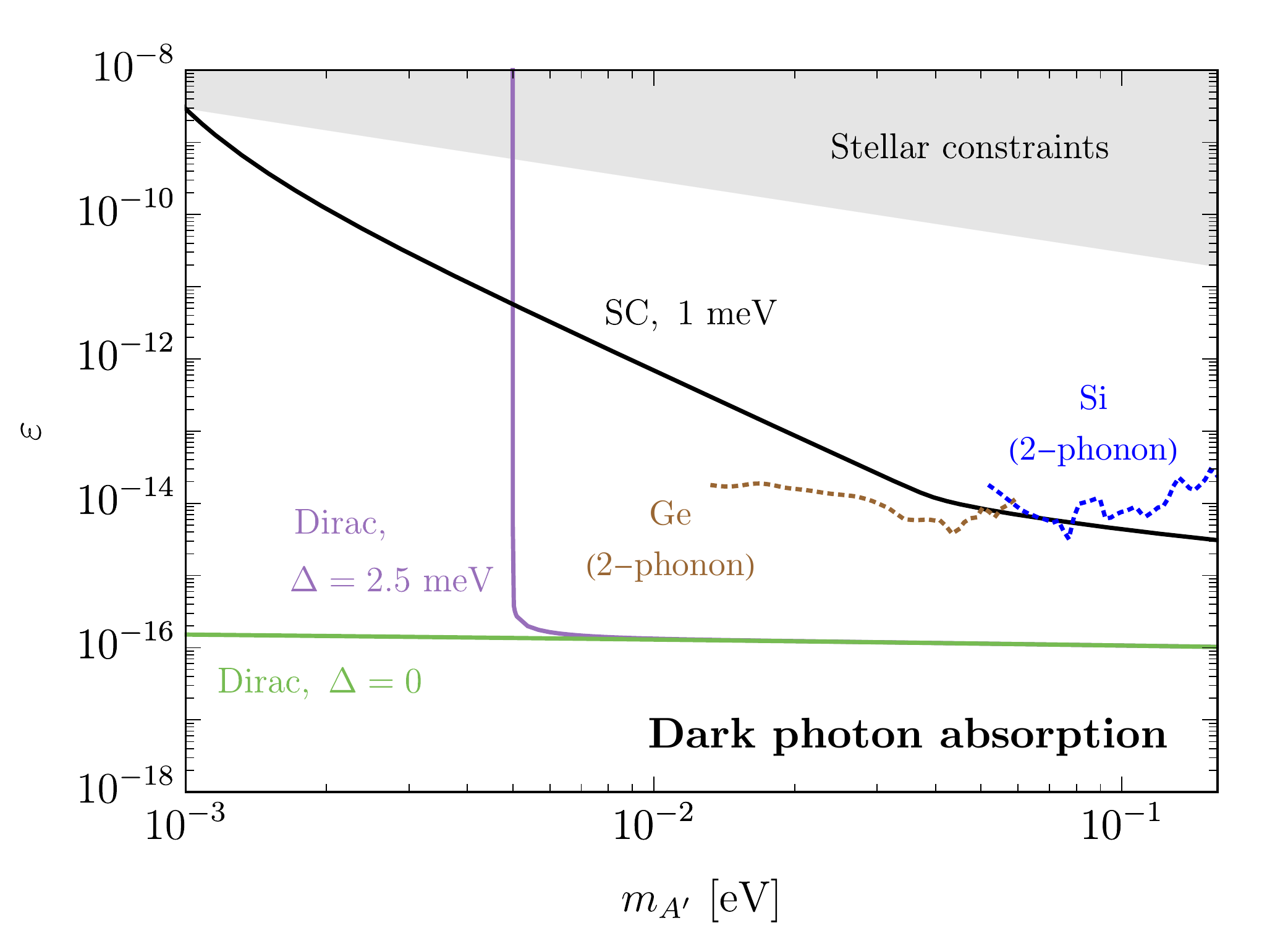}
\caption{ \label{fig:SMabsorption}
Projected reach for absorption of kinetically mixed dark photons, given in terms of the kinetic mixing parameter $\varepsilon$. We show the expected background-free 95\% C.L. sensitivity (3.0 events) that can be obtained with 1 kg-yr exposure.  The green (purple) curves are gapless (gapped) isotropic Dirac materials with $\rho_T=10$~g/cm$^3$ and all other parameters as in Fig.~\ref{fig:SMkinmix}.  We cut off the plot at $m_{A'} = 2 \Lambda v_F = 160 \ \meV$, the largest energy deposit consistent with the linear dispersion relation with cutoff $\Lambda = 0.2 \ \keV$. For comparison we show the projected reach of superconductors with a 1~meV threshold~\cite{Hochberg:2016ajh} (black), as well as two-phonon excitations in germanium (brown) and silicon (blue) semiconductors~\cite{Hochberg:2016sqx}. Stellar emission constraints~\cite{An:2014twa,An:2013b} are shown in shaded gray. }
\end{center}
\end{figure*}

The projected sensitivity of Dirac materials to absorption of a kinetically mixed dark photon is shown in  Fig.~\ref{fig:SMabsorption}, assuming 1 kg-year exposure and that the dark photon comprises all of the DM. Here, we use a typical target mass density of $\rho_T=10$~g/cm$^3$, with all other parameters equal to the fiducial parameters taken in Section~\ref{ssec:projections}. The green (purple) curves correspond to ideal isotropic gapless (gapped) systems. We do not show the projected reach for our candidate target material ZrTe$_5$, as it is highly anisotropic, not only in its band structure but also in its background dielectric tensor.  Because the kinematics of absorption dictates that $\Pi_L$ and $\Pi_T$ are of the same order of magnitude, they can mix in a potentially non-trivial way in an anisotropic medium, implying that the formalism to describe the absorption rate will be more involved.  This feature could give rise to interesting directional dependence in the rate, which will require a dedicated analysis; we leave this for future work.

For comparison we show the reach of superconductors as well as multiphonon excitations in germanium and silicon semiconductors, as obtained in Refs.~\cite{Hochberg:2016ajh,Hochberg:2016sqx}. Stellar emission constraints~\cite{An:2014twa,An:2013b} are shown in shaded gray. Note that we do not show the projected reach for magnetic bubble chambers~\cite{Bunting:2017net} as they cannot be directly compared without full treatment of in-medium effects in those systems. We find that as in the case of scattering, in-medium effects suppress the response of superconductors compared to Dirac materials. We learn that Dirac materials are excellent target materials for absorption of dark photon DM, with projected reach exceeding all current proposals when $2\Delta < m_{A'}$.

In Appendix~\ref{app:ALPs}, we discuss the reach for axion-like particle (ALP) DM, where in-medium effects are absent. As expected, because of the lower phase space volume of target electrons in a Dirac material in comparison to metals, the reach of an aluminum superconductor is superior to that of a Dirac material for ALP DM.

\section{\textbf{Conclusions}}
\label{sec:conclusions}

In this paper, we have shown that 3D Dirac materials are excellent targets to use in searches for sub-MeV DM. We have described their interactions with a kinetically mixed dark photon, and in particular the effect of the complex permittivity on the in-medium dark photon couplings.  We find that the dark photon does not develop an in-medium effective mass in these materials.  This result allows Dirac materials to probe both DM scattering and absorption involving a light kinetically mixed dark photon with greater sensitivity than any other proposed experimental target. In the case of DM scattering via a light kinetically mixed dark photon, the reach is several orders of magnitude stronger than that required to probe the theoretical benchmark of freeze-in DM, even for realistic materials. We have identified promising Dirac material candidates, including ZrTe$_5$, and determined that Fermi velocities of order $\sim10^{-4}$ or smaller are optimal for both scattering and absorption.

The strong dependence of the projected scattering reach on the Fermi velocity offers interesting possibilities for probing the DM velocity distribution. Since low-energy scattering is kinematically forbidden if $v_{\rm DM} < v_F$, repeating the same experiment with two different materials that have different Fermi velocities, one well below and one well above the maximum DM velocity, should result in marked differences in the total scattering rate.\footnote{Dirac materials may offer an interesting opportunity to probe a fast sub-population of DM reflected from the sun \cite{An:2017ojc}.} This could also serve to reduce backgrounds since inter-band excitations from slow-moving neutrons or alpha particles are always forbidden, providing a handle on backgrounds that can dominate in other experimental targets. Furthermore, anisotropic crystal lattices have been shown to have excellent directional-detection potential \cite{Hochberg:2016ntt,Budnik:2017sbu}, and the combined anisotropies of the crystal lattice and the Fermi velocities in materials such as ZrTe$_5$ suggest a similar advantage for Dirac materials. We leave an analysis of the directional-detection capabilities of anisotropic Dirac materials, for both scattering and absorption, to future work.

As with any new detection technology, many hurdles must be overcome to translate sensitivity estimates into a feasible experimental implementation. Detection of single-electron excitations in semiconductors is a burgeoning field (see Ref.~\cite{Battaglieri:2017aum} for a recent review), but the threshold energy (due to the band gap) tends to be at least $\sim 1 \mbox{ eV}$, much higher than what we consider here.  Detection of meV athermal phonons and quasiparticles has been proposed in Refs.~\cite{Hochberg:2015pha,Hochberg:2015fth}, where the detection scheme takes advantage of long excitation lifetimes made possible by ultra-pure materials such as superconducting aluminum. In recent years, many semimetal candidates have been produced and are being discovered in the laboratory~\cite{AMV17}, which increases the likelihood of finding ideal materials for DM detection. While the focus is not yet on mass fabrication techniques, ultra-pure semimetals with $\sim$mm carrier mean free paths have recently been synthesized~\cite{1703.04527}.  Moreover, since the spectrum of DM-induced excitations is peaked away from zero, some remnants of the initial excitation process may survive after relaxation. Indeed, in undoped graphene, intra-band de-excitation is highly suppressed and carrier multiplication may be the dominant relaxation process~\cite{winzer2010carrier,winzer2012impact}. It would be interesting to investigate whether the same holds true for 3D Dirac materials. In a forthcoming paper, we plan to consider these issues in depth and present a detailed experimental configuration for detection of meV-scale DM-induced excitations in Dirac materials.

Sub-MeV dark matter is a viable theoretical and experimental possibility, posing interesting challenges to both theory and experiment. The DM direct detection community is pursuing a robust suite of approaches for sub-GeV DM~\cite{Battaglieri:2017aum}, and we look forward to Dirac materials joining the hunt.

\mysection{Acknowledgements}
We thank Jens H. Bardarson, Ilya Belopolski, Rouven Essig, Snir Gazit, Zahid Hasan, Pablo Jarillo-Herrero, and Zohar Ringel for useful conversations. YH is supported by the U.S.\ National Science Foundation, grant NSF-PHY-1419008, the LHC Theory Initiative. ML is supported by the DOE under contract DESC0007968, the Alfred P.~Sloan Foundation and the Cottrell Scholar Program through the Research Corporation for Science Advancement.  KZ is supported by the DOE under contract DE-AC02-05CH11231. AGG was supported by the Marie Curie Programme under EC Grant Agreement No.\ 653846. SMG, ZL, SFW and JBN are supported by the Laboratory Directed Research and Development Program at the Lawrence Berkeley National Laboratory under Contract No.\ DE-AC02-05CH11231. This work is also supported by the Molecular Foundry through the DOE, Office of Basic Energy Sciences under the same contract number. SFW is supported in part by an NDSEG fellowship. This research used resources of the National Energy Research Scientific Computing Center, which is supported by the Office of Science of the U.S.\ Department of Energy. This work was performed in part at the Aspen Center for Physics, which is supported by National Science Foundation grant PHY-1607611. 

\appendix
%

\section{Transition Form Factor}
\label{app:Overlap}

This appendix includes the details behind the analytical form of the transition form factor Eq.~\eqref{eq:TFFAnalytic}, the relation between the transition form factor and the $\vecq^2$ dependence of the complex permittivity, and the relation to previous work \cite{Essig:2015cda} on electron scatterings with generic band structure.
\subsection{Derivation of the Transition Form Factor}
The Dirac Hamiltonian Eq.~\eqref{eq:DiracHam} in a block off-diagonal form reads
\begin{equation}
H_{\vecl} =
\left(
\begin{array}{cc}
0 & \widetilde{\vecl} \cdot\boldsymbol{\sigma} -i \Delta  \\
 \widetilde{\vecl} \cdot\boldsymbol{\sigma} +i \Delta &0
\end{array}
\right) \, ,
\end{equation}
where an anisotropic Fermi velocity is allowed by defining the rescaled momentum
\be
\widetilde{\vecl} = (v_{F,x}\ell_x,v_{F,y}\ell_y,v_{F,z}\ell_z).
\label{eq:rescalemom}
\ee
The normalized eigenstates can be written as
\begin{equation}
\label{eq:wfcDirac}
u_{1,3}^{\vecl} = \dfrac{1}{\sqrt{2}E_{\vecl}}
\left(
\begin{array}{c}
\mp \lt_{-} \\
\pm(i \Delta + \lt_z)\\
0\\
E_{\vecl}
\end{array}
\right),
\hspace{1cm}
u_{2,4}^{\vecl} = \dfrac{1}{\sqrt{2}E_{\vecl}}
\left(
\begin{array}{c}
 \pm(i \Delta - \lt_z)\\
 \mp \lt_{+}\\
E_{\vecl}\\
0
\end{array}
\right),
\end{equation}
where $\lt_{\pm}=\lt_x \pm i \lt_y$ and $E_{\vecl}=\sqrt{\veclt^2+\Delta^2}$. The upper (lower)
signs correspond to negative (positive) energy solutions $E^{\lambda}_{\vecl}=  \lambda E_{\vecl}$ where $\lambda = \mp 1$.

The transition form factor appears in the expression for the polarization function, which can be written
\beq
\Pi(\omega,\vecq) = \lim_{\eta \to 0} \frac{g}{V}\int \frac{d^3 \vecl}{(2\pi)^3} \sum_{\lambda,\lambda'} \frac{f_{\rm FD}(E^{\lambda'}_{\vecl + \vecq}) - f_{\rm FD}(E^\lambda_{\vecl})}{E^{\lambda'}_{\vecl + \vecq} - E^{\lambda}_{\vecl} - \omega - i \eta} |f_{\lambda,\vecl \to \lambda',\vecl + \vecq}|^2 \,.
\label{eq:polfunc}
\eeq
Here, $V$ is the crystal volume, $g = 2$ is the spin degeneracy, and $f_{\rm FD}$ is the Fermi-Dirac distribution, which is just a step function at zero temperature. The polarization function can also be defined as the product of two Green's functions \cite{AB70,throckmorton2015many}, and calculating the polarization function using Green's functions and matching to the form of \Eref{eq:polfunc} allows one to extract the transition form factor. In the diagonal basis, the Green's function is given by
\begin{equation}
\label{eq:Greensf}
G_{\omega,\vecl}=\sum_{\lambda=\pm}\dfrac{1}{i\omega- E^{\lambda}_{\vecl}}P^{\lambda}_{\vecl}
\end{equation}
in terms of the projection operator:
\begin{equation}
P^{\lambda}_{\vecl} =\left. | \vecl,\lambda \right\rangle \left\langle  \vecl, \lambda \right | =
 \dfrac{1}{2}\left(
\begin{array}{cc}
1 & \frac{\lambda }{E_{\vecl}} \left(\veclt\cdot\boldsymbol{\sigma} - i \Delta \right)  \\
\frac{\lambda }{E_{\vecl}} \left(\veclt\cdot\boldsymbol{\sigma}+ i \Delta \right) &1
\end{array}
\right),
\end{equation}
formally defined using the eigenvectors Eqs.~\eqref{eq:wfcDirac}, {\it i.e.} $P^{+}_{\vecl} =  \left. | \vecl,+ \right\rangle \left\langle  \vecl,+ \right |
=  \sum_{i=1,2} \left | u_{i}^\vecl  \right\rangle \left\langle u_{i}^{\vecl}  \right |$ with an analogous definition for $P^{-}_{\vecl} $.
As is standard in many-body physics~\cite{AB70,throckmorton2015many}, the polarization function results after integrating over the frequency domain and taking the trace of the product of two Green's functions given by Eq.~\eqref{eq:Greensf} evaluated at different momenta $\vecl$ and $\vecl' = \vecl + \vecq$.
After evaluating the frequency integral, we are left with the explicit kernel of Eq.~\eqref{eq:polfunc}, a factor of $1/2$ and
the trace over the product of two projectors. The product of the latter two objects defines the transition form factor as
\begin{eqnarray}
\label{eq:Tr}
 |f_{\lambda,\vecl \to \lambda',\vecl'}|^2 &=& \dfrac{1}{2}\mathrm{Tr}\left[P^{\lambda}_{\vecl}P^{\lambda'}_{\vecl'}\right] = \dfrac{1}{2} \mathrm{Tr}\left[ | \vecl,\lambda\right\rangle
\left\langle \vecl,\lambda || \vecl',\lambda'  \right\rangle   \left\langle  \vecl',\lambda' |  \right]\nonumber \\
&=& \dfrac{1}{2}\left(1+ \frac{\lambda \lambda'}{E_{\vecl}E_{\vecl'}} (\veclt \cdot \veclt'+\Delta^2)\right).
\label{eq:TFFGap}
\end{eqnarray}

It is worth noting that the overlap factor for Dirac systems can be related to standard completeness relations over spinors, which are perhaps more familiar in the high-energy theory context.

\subsection{Relation Between Permittivity and Transition Form Factor}
The complex permittivity of a material is given in general by the Lindhard formula~\cite{dressel}:
\beq
\epsilon_r(\omega,\vecq) = 1 - \lim_{\eta \to 0} \frac{1}{V}\frac{e^2}{\kappa}\frac{1}{\vecq^2} \int_{\rm BZ} \frac{d^3 \veck}{(2\pi)^3} \sum_{n,n'} g_{s,n} \frac{f_{\rm FD}(E_{\veck + \vecq, n'}) - f_{\rm FD}(E_{\veck,n})}{E_{\veck + \vecq, n'} - E_{\veck,n} - \omega - i \eta} |f_{n,\veck \to n',\veck + \vecq}|^2 \, .
\label{eq:Lindhard}
\eeq
Here, $\kappa$ is the background dielectric constant, $n$ and $n'$ are band indices, $g_{s,n}$ is the spin degeneracy of band $n$, $E_{\veck,n}$ is the energy of the $n^\text{th}$ band at lattice momentum $\veck$, and the integral is taken over the first Brillouin zone (BZ).  The Fermi-Dirac factors in the numerator ensure that at zero temperature, the only transitions which contribute to $\epsilon_r$ are from unoccupied to occupied states and vice-versa. Using \Eref{eq:TFFGap} and performing the momentum integral in Eq.~\eref{eq:Lindhard} with a cutoff $\Lambda$, considering a single Dirac cone with valence and conduction bands $n = -$ and $n' = +$ only, yields the dielectric constant for an ideal Dirac material. This is equivalent to evaluating \Eref{eq:polfunc} and using the relationship between the polarization function and the permittivity \cite{dressel}. The integral can be performed analytically for $\Delta = 0$, yielding \Eref{eq:epsSM}, which we repeat here for convenience:

\beq\label{eq:epsSMApp}
\left(\epsilon_r\right)_{\rm semimetal}=1-\frac{e^2g}{24\pi^2 \kappa v_F}\frac{1} {{\bf q}^2}\left\{-\vecq ^2{\rm ln}\left|\frac{4\Lambda^2}{\omega^2/v_F^2-\vecq^2}\right|-i\pi \vecq^2 \Theta(\omega-v_F |{\bf q}|)
\right\}\,.
\eeq
We have written Eq.~\eqref{eq:epsSMApp} in a form resembling Eq.~\eqref{eq:Lindhard} to illustrate that the inter-band transitions yield a form factor that scales as $\vecq^2$ (to leading order), which multiplies the Fourier transform of the Coulomb potential $e^2/(\kappa \vecq^2)$. This behavior is a direct consequence of the orthogonality of the valence and conduction bands, which implies that the transition form factor \Eref{eq:fgen} vanishes at $\vecq = 0$. This remains true for nonzero $\Delta$. Indeed, defining $\widetilde{\Delta} = \Delta/v_F$ and expanding \Eref{eq:TFFGap} in small $\vecq$ yields
\be
|f_{-,\vecl \to +,\vecl+\vecq}(\mathbf{q})|^2 = \frac{1}{4}\left(\frac{\vecq^2}{\vecl^2+\widetilde{\Delta}^2} - \frac{(\vecl \cdot \vecq)^2}{(\vecl^2+\widetilde{\Delta}^2)^2}\right) + \mathcal{O}(\vecq^4)\,.
\ee

This derivation illustrates an alternative perspective on why the dark photon does not develop an effective mass in-medium in a Dirac material: the vanishing of $|f_{-,\vecl \to +,\vecl+\vecq}|^2$ as $\vecq \to 0$ in \Eref{eq:Lindhard} ensures that $\epsilon_r$ is constant as $\vecq \to 0$, or equivalently $\Pi(q) \sim q^2$.

\subsection{Lattice Momentum Conservation and Comparison to Formalism For Generic Band Structure}
\label{ssec:latticemom}

Note that \Eref{eq:Tr} is written only as a function of initial and final state momenta. To make contact with the formalism of Ref.~\cite{Essig:2015cda}, we will also write it as a function of the momentum transfer $\vecq$ by inserting unity in the form $\frac{(2\pi)^3}{V}\delta(\vecq - (\vecl' - \vecl))$, where $V$ is the volume of the crystal:
\be
\label{eq:TFFqform}
 |f_{\lambda,\vecl \to \lambda',\vecl'}(\vecq)|^2 =  \frac{(2\pi)^3}{V} \delta(\vecq - (\vecl' - \vecl)) \frac{1}{2}\left(1+ \lambda \lambda'\, \frac{\veclt \cdot \veclt'+\Delta^2}{\sqrt{\veclt^2 + \Delta^2}\sqrt{\veclt'^2 + \Delta^2}} \right),
 \ee
 which reduces to \Eref{eq:TFFAnalytic} in the gapless isotropic limit.

The delta function in \Eref{eq:TFFqform} enforces exact momentum conservation, while typically in problems involving condensed matter systems, momentum is only conserved up to addition of a reciprocal lattice vector. We can justify the exact momentum conservation for the cases relevant to DM scattering by using kinematic arguments. A convenient parameterization of a general wavefunction in a periodic system is as a linear combination of Bloch waves,
\be
\Psi_{n, \mathbf{k}}( \mathbf{ x}) = \frac{1}{\sqrt{V}}\sum_{\vecG} u_n(\mathbf{k}+\mathbf{G})e^{i (\mathbf{k}+\mathbf{G})\cdot \mathbf{x}}\,,
\label{eq:BlochExpansion}
\ee
where $\vecG$ runs over all reciprocal lattice vectors and the $u_n$ are complex numbers. The velocity- and directionally-averaged scattering rate for a single electron in the Bloch basis is \cite{Essig:2015cda}
\be
R_{n,\veck \to n',\veck'} = \left .\frac{\rho_\chi}{m_\chi} \frac{\pi^2 \bar{\sigma}_e} { \mu_{\chi e}^2}\frac{1}{V} \sum_{\vecG'}\frac{1}{|\vecq|} \eta\left(v_{\rm min}(|\vecq|,  \omega_{\veck \veck'})\right)|F_{\rm DM}(q)|^2 |f_{[n \veck,n'\veck',\vecG']}|^2 \right |_{\vecq = \veck' - \veck + \vecG'}
\label{eq:Rik},
\ee
where the crystal form factor is
\be
f_{[n \veck,n'\veck',\vecG']} = \sum_{\vecG} u^*_{n'}(\veck' + \vecG + \vecG') \, u_{n}(\veck + \vecG),
\label{eq:ffactor}
\ee
which is related to the transition form factor as defined in \Eref{eq:fgen} by
\be
|f_{n,\veck \to n',\veck'}(\vecq)|^2 = \sum_{\vecG'}\frac{(2\pi)^3}{V}\delta(\vecq - (\veck' -\veck + \vecG'))|f_{[n \veck,n'\veck',\vecG']}|^2.
\label{eq:frelation}
\ee

In the Bloch wave basis, orthogonality of different Fourier components leads to lattice momentum conservation $\vecq = \veck' - \veck + \vecG'$. Here, $\vecG'$ is a multiple of a reciprocal lattice vector whose size $|\vecG'|$ is either 0 or $\sim 2\pi m/a$ for integers $m >0$, where $a$ is the lattice spacing. A typical lattice will have $a \sim 1$--$10~\ang$, so $2\pi/a \sim \keV$.

If $m \neq 0$ we have $|\vecG'| \gtrsim \keV$. In a Dirac material, transitions near the Dirac point satisfy $\veck' - \veck = \vecl' - \vecl$ with $|\vecl|, |\vecl'| \ll |\vecG'|$ by assumption. Thus $|\vecq| \sim |\vecG'| \gtrsim \keV$. Referring to Eq.~(\ref{eq:vmin}), $v_{\rm min} \geq \frac{|\vecq|}{2m_\chi} \gtrsim 10^{-2} \gg v_{\rm DM}$ for $m_\chi \lesssim 100 \ \keV$. In other words, scattering is kinematically impossible for $m_\chi \lesssim 100 \ \keV$ unless $\vecG' = 0$. Even if scattering is kinematically allowed for $\vecG' \neq 0$, we will be primarily concerned with form factors which scale as $F_{\rm DM}(q) \sim 1/q^2$, so that the rate \eqref{eq:Rik} scales as $1/|\vecq|^5$. This represents an enormous suppression when $\vecG' \neq 0$ of the order of $({\rm eV}/\keV)^5 \simeq 10^{-15}$. Thus in our kinematic regime, reciprocal vectors $\vecG' \neq 0$ can be safely neglected.

When $\vecG' = 0$, the sum in \Eref{eq:frelation} collapses to a single term, and we can identify
\be
|f_{[(n=-) \veck,(n'=+) \veck',0]}|^2 = \frac{1}{2}\left(1+ \lambda \lambda'\, \frac{\veclt \cdot \veclt'+\Delta^2}{\sqrt{\veclt^2 + \Delta^2}\sqrt{\veclt'^2 + \Delta^2}} \right).
\label{eq:Gzero}
\ee
The single delta function in \Eref{eq:frelation} now enforces $\vecq = \vecl' - \vecl$, establishing that in our kinematic regime, the physical momentum transfer $\vecq$ is equal to the difference in lattice momenta between initial and final states.

\section{Modifications for Anisotropic Dirac Materials}
\label{app:aniso}

In this Appendix we discuss modifications to our analysis for scattering of light DM in Dirac materials for the case of anisotropic materials, with $v_{F,x} \neq v_{F,y} \neq v_{F,z}$.

\subsection{Anisotropic Permittivity}
\label{ssec:anperm}

For anisotropic Dirac materials, one may make a change of variables in the integrand of Eq.~\eqref{eq:Lindhard} and evaluate the permittivity at a correspondingly rescaled value of the momentum \cite{PhysRevD.86.045001,1501.04636}:
\be
\left(\epsilon_r\right)_{\rm Dirac}^{\rm an.} = 1 - \frac{\widetilde{\vecq}^2}{\vecq^2}\frac{1 }{v_{F,x} v_{F,y} v_{F,z}} \left(1 - \epsilon_r^{\rm iso.}(\widetilde{\vecq})|_{v_F = 1}\right).
\ee
Here, $\widetilde{\vecq}$ is defined as in \Eref{eq:rescalemom}, $\widetilde{\vecq} = (v_{F,x}q_x,v_{F,y}q_y,v_{F,z}q_z)$, and on the right-hand side the isotropic form factor is evaluated for $v_F = 1$ and at the rescaled momentum $\widetilde{\vecq}$. For example, in the gapless case, Eq.~\eref{eq:epsSM} is modified to
\be
\left(\epsilon_r\right)_{\rm semimetal}^{\rm an.} = 1 - \frac{1}{ \vecq^2}\frac{e^2g}{24\pi^2\kappa v_{F,x} v_{F,y} v_{F,z}} \left\{-\widetilde{\vecq}^2 \ln \left|\frac{4\widetilde{\Lambda}^2}{\omega^2 - \widetilde{\vecq}^2}\right|- i\pi\widetilde{\vecq}^2\Theta(\omega-|\widetilde{\vecq}|)\right\}\,,
\label{eq:epsAn}
\ee
The cutoff $\Lambda$ must also be rescaled: we choose $\widetilde{\Lambda} = \Lambda \times {\rm max}(v_{F,x},v_{F,y},v_{F,z}) $ rather than {\em e.g.}\ $ \Lambda \times (v_{F,x} v_{F,y} v_{F,z})^{1/3}$ to recover the correct scaling when one of the $v_{F,i}$ is much smaller than the other two, as is typically the case with real materials.

In addition to the anisotropy of the band structure, the crystal lattice itself may be anisotropic, in which case $\epsilon_r$ is more properly described by a full tensor $(\epsilon_r)_{ij}$. In this situation, \Eref{eq:Lindhard} should be interpreted as a tensor equation. In the basis of principal components where the background dielectric tensor $\kappa_{ij}$ is diagonal, $(\epsilon_r)_{ij}$ is also diagonal. In the gapless case its diagonal components are given by
\be
\left(\epsilon_r\right)_{ii} = 1 - \frac{1}{ \vecq^2}\frac{e^2g}{24\pi^2\kappa_{ii} \, v_{F,x} v_{F,y} v_{F,z}} \left\{-\widetilde{\vecq}^2 \ln \left|\frac{4\widetilde{\Lambda}^2}{\omega^2 - \widetilde{\vecq}^2}\right|- i\pi\widetilde{\vecq}^2\Theta(\omega-|\widetilde{\vecq}|)\right\}\,,
\ee
with straightforward modifications for the gapped case. Strictly speaking, the formalism of Section~\ref{sec:formalism} does not apply because longitudinal and transverse modes are not decoupled in anisotropic media. However, for the case of scattering, $\Pi_L \gg \Pi_T$ and the dominant effects are still given by $\Pi_L$. Assuming a spherically-symmetric velocity distribution, the leading effects of the anisotropic tensor $(\epsilon_r)_{ij}$ can be captured by its rotationally-invariant component $\frac{1}{3}\Tr(\epsilon_r)$:
\be
\frac{1}{3}\Tr \left(\epsilon_r\right)= 1 - \frac{1}{ \vecq^2}\frac{e^2g}{24\pi^2\widetilde{\kappa} \, v_{F,x} v_{F,y} v_{F,z}} \left\{-\widetilde{\vecq}^2 \ln \left|\frac{4\widetilde{\Lambda}^2}{\omega^2 - \widetilde{\vecq}^2}\right|- i\pi\widetilde{\vecq}^2\Theta(\omega-|\widetilde{\vecq}|)\right\}\,,
\ee
where
\be
\widetilde{\kappa} = \frac{3}{1/\kappa_{xx} + 1/\kappa_{yy} + 1/\kappa_{zz}}.
\ee
Therefore, in our analysis of scattering in ZrTe$_5$ where $\kappa_{ij}$ is anisotropic, we compute spherically-symmetric rates using $\widetilde{\kappa}$.

\subsection{Scattering in Anisotropic Dirac Materials}

The impact of anisotropic dispersions on the DM scattering rate, \Erefs{eq:RdiffNoGap}{eq:RtotNoGap}, can be estimated in a straightforward manner. In typical Dirac materials, the anisotropy of the Dirac cones often involves a hierarchy of Fermi velocities $v_{F,z} \ll v_{F,x}, v_{F,y}$, where $v_{F,x} \simeq v_{F,y} \equiv v_{F,\perp}$. In this limit, Eq.~(\ref{eq:epsAn}) becomes
\be
\epsilon_r^{v_F \ll v_\perp} \approx 1 - \frac{{\vecq_\perp}^2}{\kappa \vecq^2}\frac{e^2}{v_{F,z}} \left\{-\frac{g}{24\pi^2}{\rm ln}\left|\frac{4\Lambda ^2}{\omega^2/v_{F,\perp} - \vecq_\perp^2}\right|-\frac{i g}{24\pi}\Theta(\omega-v_{F,\perp}|\vecq_\perp|)\right\}\, ,
\ee
where $\vecq_\perp = (q_x, q_y, 0)$. Following the arguments of Section~\ref{ssec:ScatteringKinematics}, the total scattering rate will then be proportional to $v_{F,z}^2$, the smallest of the Fermi velocities. However, $v_{\rm min}$, which controls the behavior of the integral as a function of the Fermi velocities, now takes the form
\begin{align}
v_{\rm min}(|\vecq|, \omega_{\vecl, \vecl + \vecq}) &= \frac{\sqrt{v_{F,\perp}^2 (\vecl + \vecq)_\perp^2 + v_{F,z}^2 (\ell_z + q_z)^2} + \sqrt{v_{F,\perp}^2 \vecl_\perp^2 + v_{F,z}^2 \ell_z^2} }{|\vecq|} + \frac{|\vecq|}{2m_\chi} \nonumber \\
& = v_{F,\perp}\frac{ |\vecl_\perp + \vecq_\perp| + |\vecl_\perp|}{|\vecq|} + \frac{|\vecq|}{2m_\chi} + \mathcal{O}\left(\frac{v_{F,z}^2}{v_{F,\perp}^2}\right).
\end{align}
Here, the argument of Section~\ref{ssec:ScatteringKinematics} that DM scattering is allowed only when $v_F < v_{\rm DM}$ fails because by taking $\vecq_\perp = 0$, $\vecl_\perp$ small and $q_z$ large, the first term can be made much smaller than $v_{F,\perp}$ and scattering is allowed even when $v_{\rm DM} < v_{F,\perp}$. On the other hand, we can obtain a lower bound for $v_{\rm min}(|\vecq|, \omega_{\vecl, \vecl + \vecq})$ by taking $v_{F,\perp} = v_{F,z}$, and by repeating the kinematic argument of Section~\ref{ssec:ScatteringKinematics}, we see that we need $v_{\rm DM} > v_{F,z}$ for scattering to occur. As a result, the behavior of the integral is also dominated by the smallest velocity $v_{F,z}$, so we expect similar results to the isotropic case.

\begin{figure*}[t!]
\begin{center}
\includegraphics[width=0.6\textwidth]{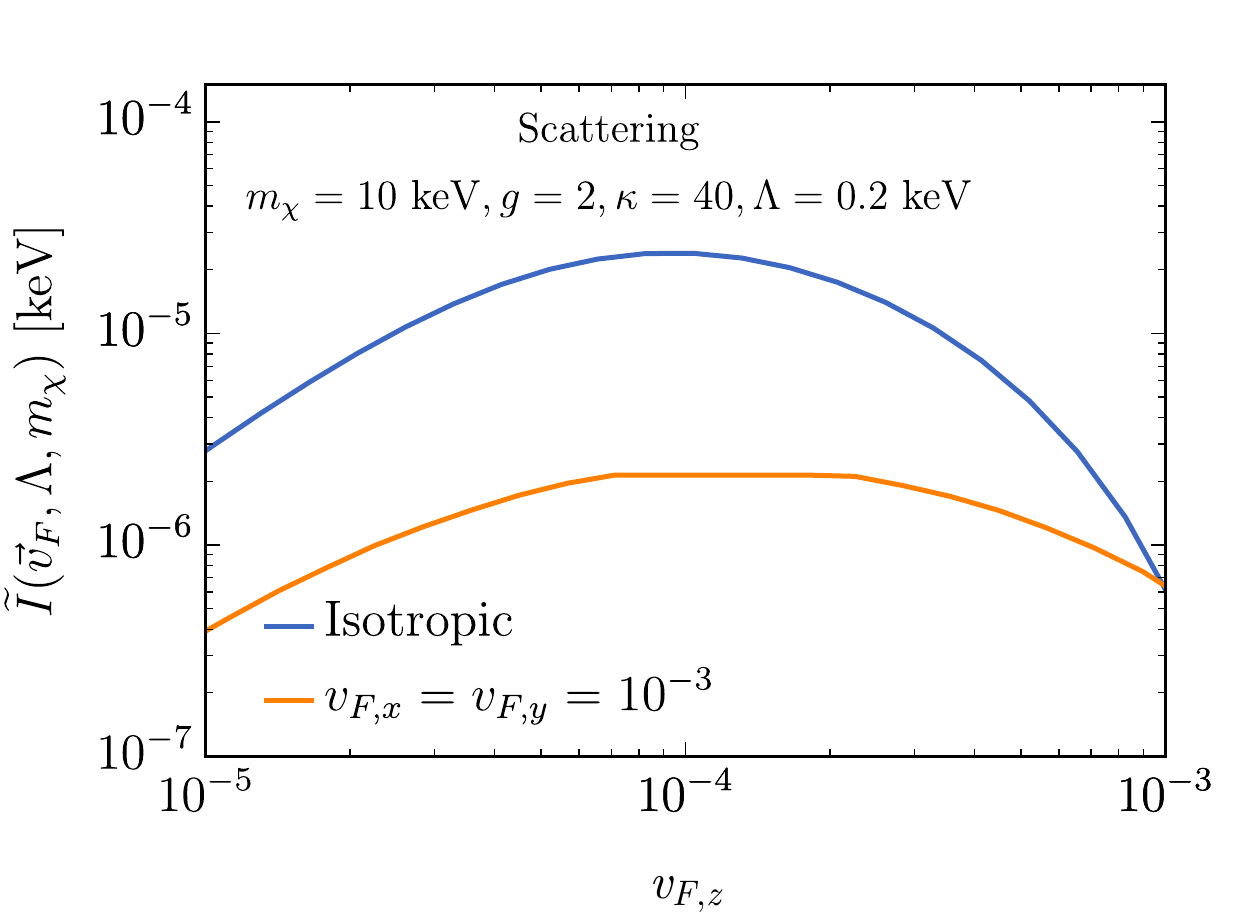}
\caption{ \label{fig:AnisotropicCompare}
Scaling of the DM scattering rate for $m_\chi = 10 \ \keV$ with the Fermi velocity $v_{F,z}$ of a gapless Dirac material, comparing an isotropic dispersion $v_{F,x} = v_{F,y} = v_{F,z}$ to one with $v_{F,x} = v_{F,y} = 10^{-3}$.}
\end{center}
\end{figure*}

To make this comparison concrete, we recall the isotropic rate integral $I(v_F, \Lambda, m_\chi)$ implicitly defined in \Eref{eq:RtotNoGap}, with explicit expression
\be
I(v_F, \Lambda, m_\chi) = \int_0^\Lambda d|\vecl| \int_{-1}^{1} d\cos \theta_{\ell q} \int_0^{q_{\rm max}} d|\vecq|  \frac{|\vecl|^2 |\vecq|}{(\omega_{\vecl, \vecl+\vecq}^2 - \vecq^2)^2}
\frac{\eta\left(v_{\rm min}(|\vecq|, \omega_{\vecl,\vecl+\vecq})\right)}{\ln^2 \left|\frac{4\Lambda'^2}{\omega_{\vecl,\vecl+\vecq}^2/v_F^2-\vecq^2}\right| + \pi^2} \left(1 - \frac{ \vecl \cdot (\vecl + \vecq)}{|\vecl| |\vecl + \vecq|}\right).
\label{eq:IvFLambdaM}
\ee
The limit of integration $q_{\rm max} = - |\vecl| \cos \theta_{\ell q} + \sqrt{\Lambda^2 - \vecl^2(1-\cos^2 \theta_{\ell q})}$ ensures $|\vecl + \vecq| < \Lambda$. We now define a generalized anisotropic rate integral
\be
\widetilde{I}(\vec{v}_F, \Lambda, m_\chi) = \frac{e^4 g^2}{2304 \pi^6} \int d^3 \vecq \, d^3 \vecl \frac{1}{|\vecq|} \frac{1}{(\omega_{\vecl, \vecl+\vecq}^2 - \vecq^2)^2}
\frac{\eta\left(v_{\rm min}(|\vecq|, \omega_{\vecl,\vecl+\vecq})\right)}{|\epsilon_r^{{\rm an.}}(\omega_{\vecl,\vecl+\vecq},\vecq)|^2} \left(1 - \frac{ \veclt \cdot (\veclt + \widetilde{\vecq})}{|\veclt| |\veclt + \widetilde{\vecq}|}\right),
\label{eq:ItildevFLambdaM}
\ee
where $\vec{v}_F = (v_{F,x},v_{F,y},v_{F,z})$, $\epsilon_r^{{\rm an.}}$ is defined in \Eref{eq:epsAn}, and the rescaled momenta $\widetilde{\vecq}, \veclt$ are defined as in \Eref{eq:rescalemom}. This is related to the isotropic rate integral (\ref{eq:IvFLambdaM}) by $\tilde{I} =  \frac{\kappa^2}{g} v_F^2 I$ in the isotropic case $\vec{v}_F = (v_F, v_F, v_F)$. In Fig.~\ref{fig:AnisotropicCompare}, we plot $\widetilde{I}(\vec{v}_F,\Lambda,m_\chi)$ for $m_\chi = 10 \ \keV$, $\Lambda = 0.2 \ \keV$, and $v_{F,x} = v_{F,y} = 10^{-3}$ as a function of $v_{F,z}$. As anticipated, the shape of the two curves is qualitatively similar for $v_{F,z} \ll v_{F,x/y}$, with both curves scaling similarly at small $v_{F,z}$. However, the rate is suppressed by about an order of magnitude in the anisotropic case, showing that isotropic Dirac materials are preferred for scattering.


\section{Reach for Other Models}

In this Appendix we consider the cases of scattering through a mediator $\phi$ with \emph{no} in-medium interactions (such as a scalar), as well as the case of a heavy kinetically mixed mediator $A'$, where `heavy' means $m_{A', \phi} \gg \keV$. We also consider the case of absorption of pseudoscalar dark matter (axion-like particles or ALPs).

\subsection{Scattering Reach for Other Mediator Models}
\label{app:OtherModels}

The form factors and fiducial cross sections for light scalar mediators, heavy scalar mediators, and heavy kinetically mixed mediators take the following form:
\begin{align}
\phi, \ {\rm light}: \ &F_{\rm DM}(q)  = \frac{q_0^2}{q^2}, \ \ \mathcal{F}_{\rm med}(q) = 1, \ \ \overline{\sigma}_e = \frac{16\pi \mu_{\chi e}^2 \epsilon^2 \alpha_{\rm EM} \alpha_D}{q_0^4} \quad (q_0^2 = (\alpha_{\rm EM} m_e)^2); \\
\phi, \ {\rm heavy}: \ &F_{\rm DM}(q)  = 1, \ \ \mathcal{F}_{\rm med}(q) = 1, \ \ \overline{\sigma}_e = \frac{16\pi \mu_{\chi e}^2 \epsilon^2 \alpha_{\rm EM} \alpha_D}{m_\phi^4}; \\
A', \ {\rm heavy}: \ &F_{\rm DM}(q)  = 1, \ \ \mathcal{F}_{\rm med}(q) = \frac{1}{\epsilon_r(q)}, \ \ \overline{\sigma}_e = \frac{16\pi \mu_{\chi e}^2 \epsilon^2 \alpha_{\rm EM} \alpha_D}{m_{A'}^4}.
\end{align}
Note that since $\epsilon_r(q)$ is roughly constant from \Eref{eq:epsSM}, the in-medium form factors for the two heavy mediators are roughly proportional, with in-medium effects providing an order-1 suppression.

\begin{figure*}[t!]
\begin{center}
\includegraphics[width=0.48\textwidth]{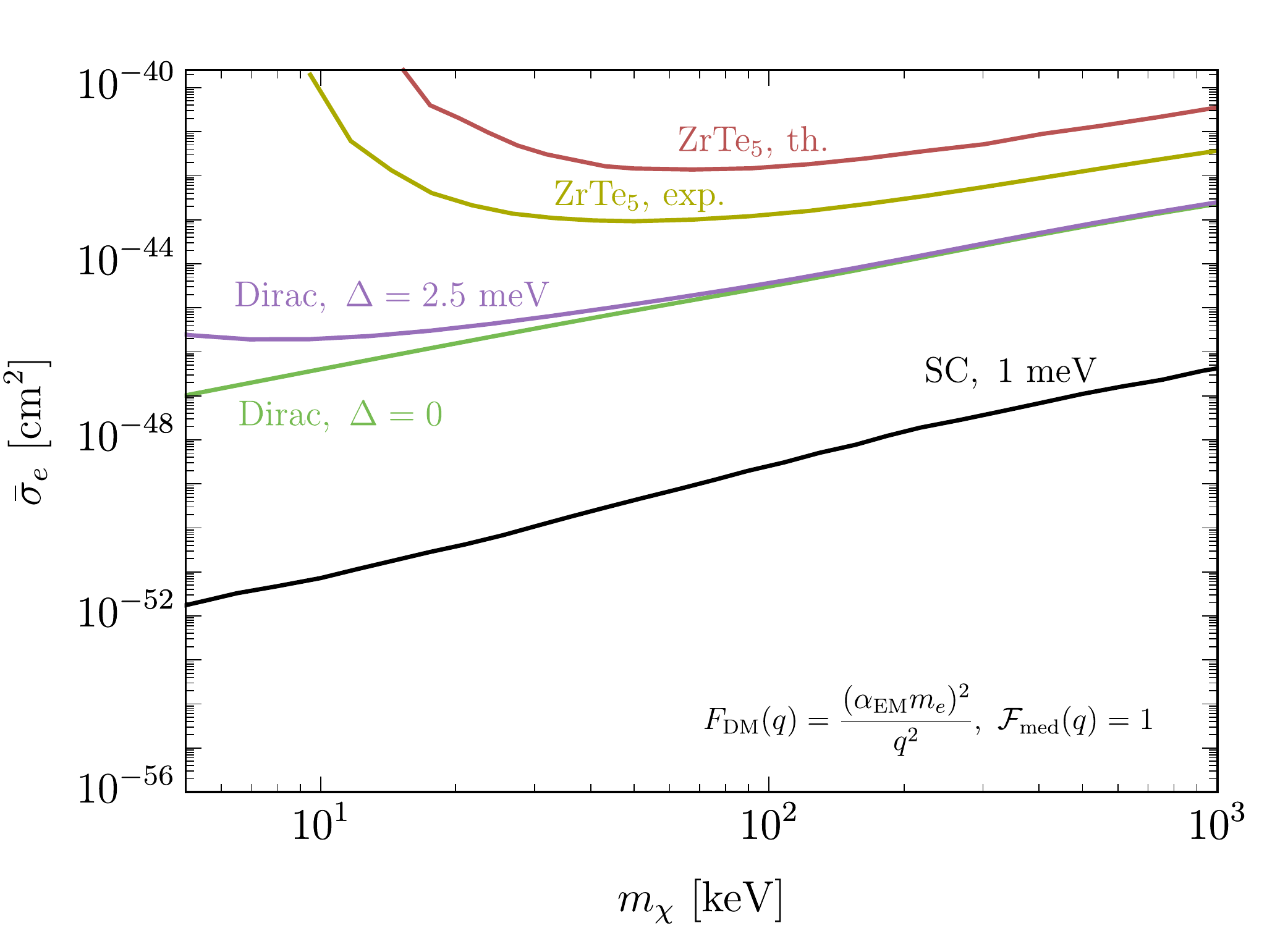}
\includegraphics[width=0.48\textwidth]{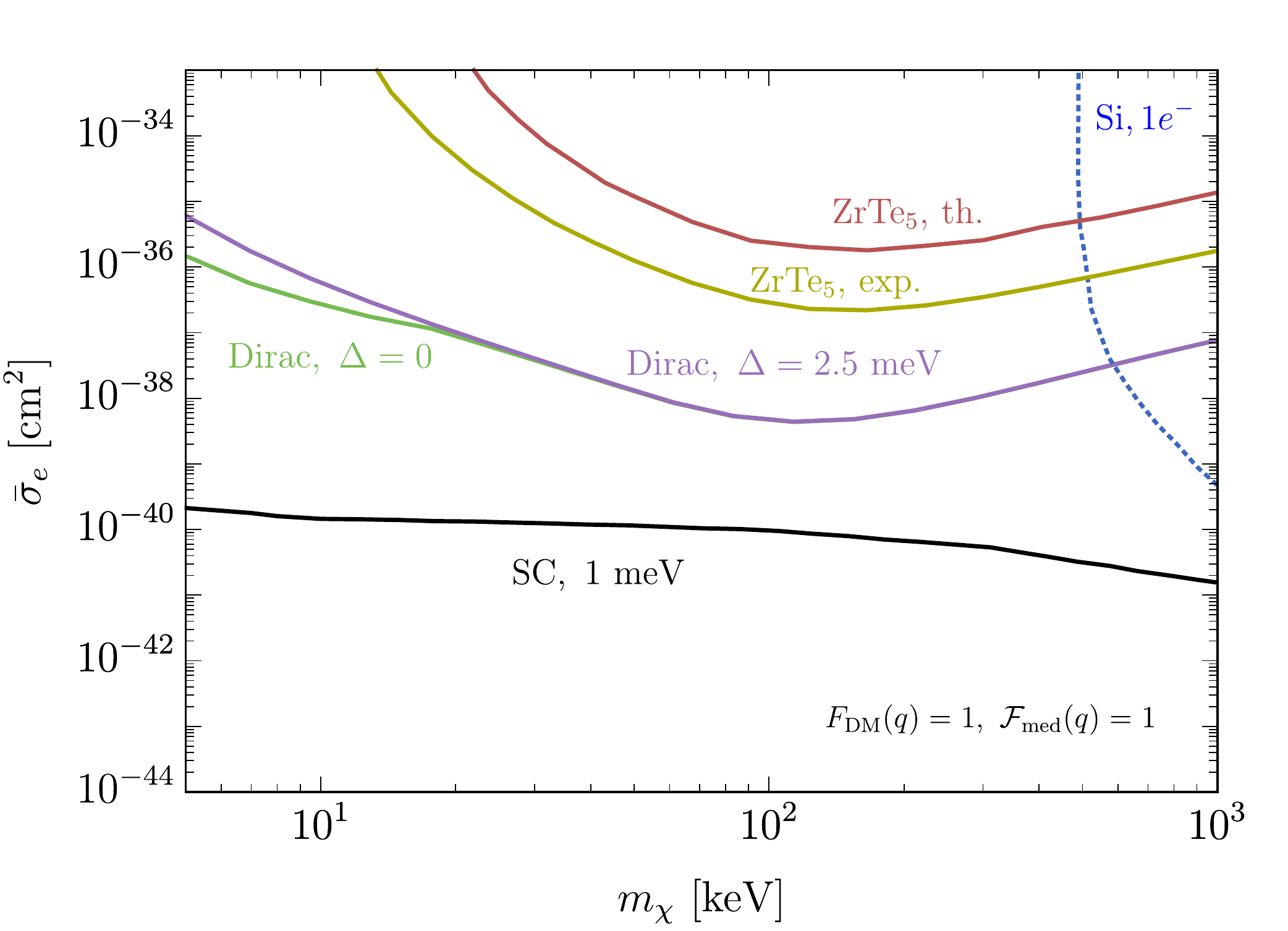}
\caption{ \label{fig:ScalarReach}
Projected scattering reach for a light (\textit{left}) and heavy (\textit{right}) mediator $\phi$ without in-medium effects. Such models are subject to strong constraints, see text for discussion. We show the expected background-free 95\% C.L. sensitivity (3.0 events) that can be obtained with 1 kg-yr exposure. Dirac material parameters are the same as in Fig.~\ref{fig:SMkinmix}.
 }
\end{center}
\end{figure*}

\begin{figure*}[t!]
\begin{center}
\includegraphics[width=0.6\textwidth]{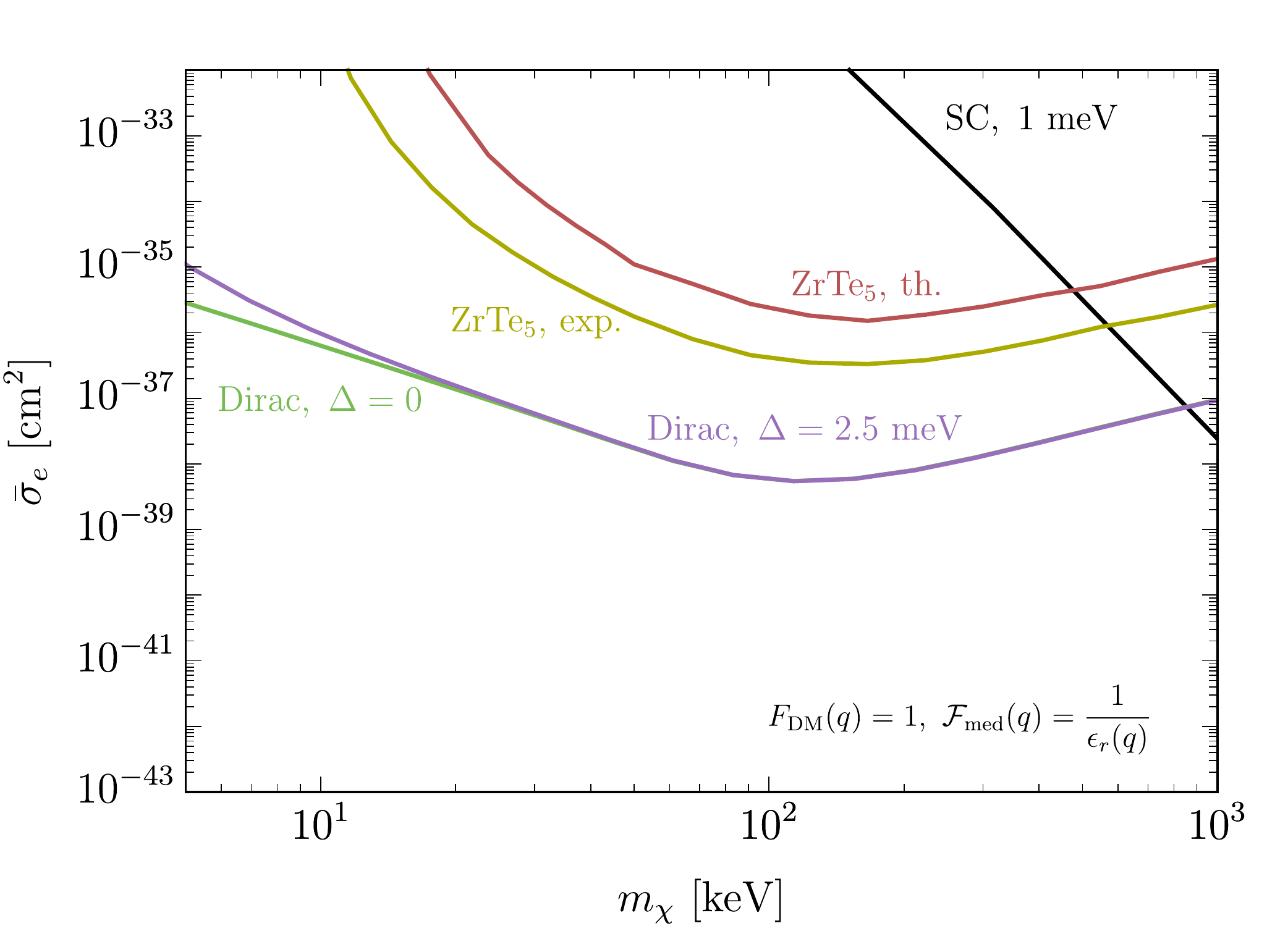}
\caption{ \label{fig:HeavyVectorReach}
Projected scattering reach for a heavy kinetically mixed mediator $A'$ including in-medium effects. Such models are subject to strong constraints, see text for discussion. We show the expected background-free 95\% C.L. sensitivity (3.0 events) that can be obtained with 1 kg-yr exposure. Dirac material parameters are the same as in Fig.~\ref{fig:SMkinmix}.
 }
\end{center}
\end{figure*}

As shown in Fig.~\ref{fig:ScalarReach}, Dirac materials have inferior reach to superconductors for mediators which are not kinetically mixed. Since in these models, the mediator does not acquire a large in-medium mass in superconductors, the larger phase space of superconductors dominates, especially for the light scalar where smaller momentum transfers are favored. 
The reach of Dirac materials compared to superconductors is slightly better for a heavy mediator than for a light mediator; the reason is that the phase space volume for a semimetal grows as $\omega^3$ compared to $\sqrt{\omega}$ for a metal, allowing much of the phase space suppression to be made up at larger energy transfers. The weakening reach of Dirac materials at masses $m_\chi \gtrsim 200 \ \keV$ for the heavy mediator is due to the phase space cutoff at $\Lambda = 0.2 \ \keV$. 
On the other hand, as shown in Fig.~\ref{fig:HeavyVectorReach}, Dirac materials have superior reach for the heavy kinetically mixed mediator, because the part of the in-medium polarization which scales as $q^2$ still suppresses the effective dark photon coupling significantly in metals.

While the DM masses and cross sections are too small to be constrained by current direct detection experiments, these models are in strong tension with astrophysical and cosmological constraints which must be evaded, at least for the most naive of models. In the massive mediator case, detectable cross-sections imply thermalization of the DM sector, including both the mediator and the DM; this, however, is in tension with Big Bang Nucleosynthesis, which requires at the $2\sigma$ level that only one real scalar, in addition to the Standard Model, can be thermalized at temperatures below an MeV \cite{Cyburt:2015mya}.\footnote{For sufficiently large cross sections, multiple scattering in the Earth may either prevent the DM from reaching the detector \cite{Emken:2017erx} or cause excessive heating in the Earth \cite{Chauhan:2016joa}. However, the relevant cross sections are much larger than those depicted in our plots.}   In the massless mediator case, stellar constraints on the emission of light mediators imply the couplings to electrons are generally too small to be detectable \cite{An:2013b}.  (The exception is a light vector particle whose mass is given by a Stueckelberg mechanism; this is the benchmark model utilized in Fig.~\ref{fig:SMkinmix}.)  These constraints are reviewed in Ref.~\cite{Hochberg:2015fth}.

\subsection{Absorption Reach for Axion-Like Particles}
\label{app:ALPs}

\begin{figure*}[t!]
\begin{center}
\includegraphics[width=0.65\textwidth]{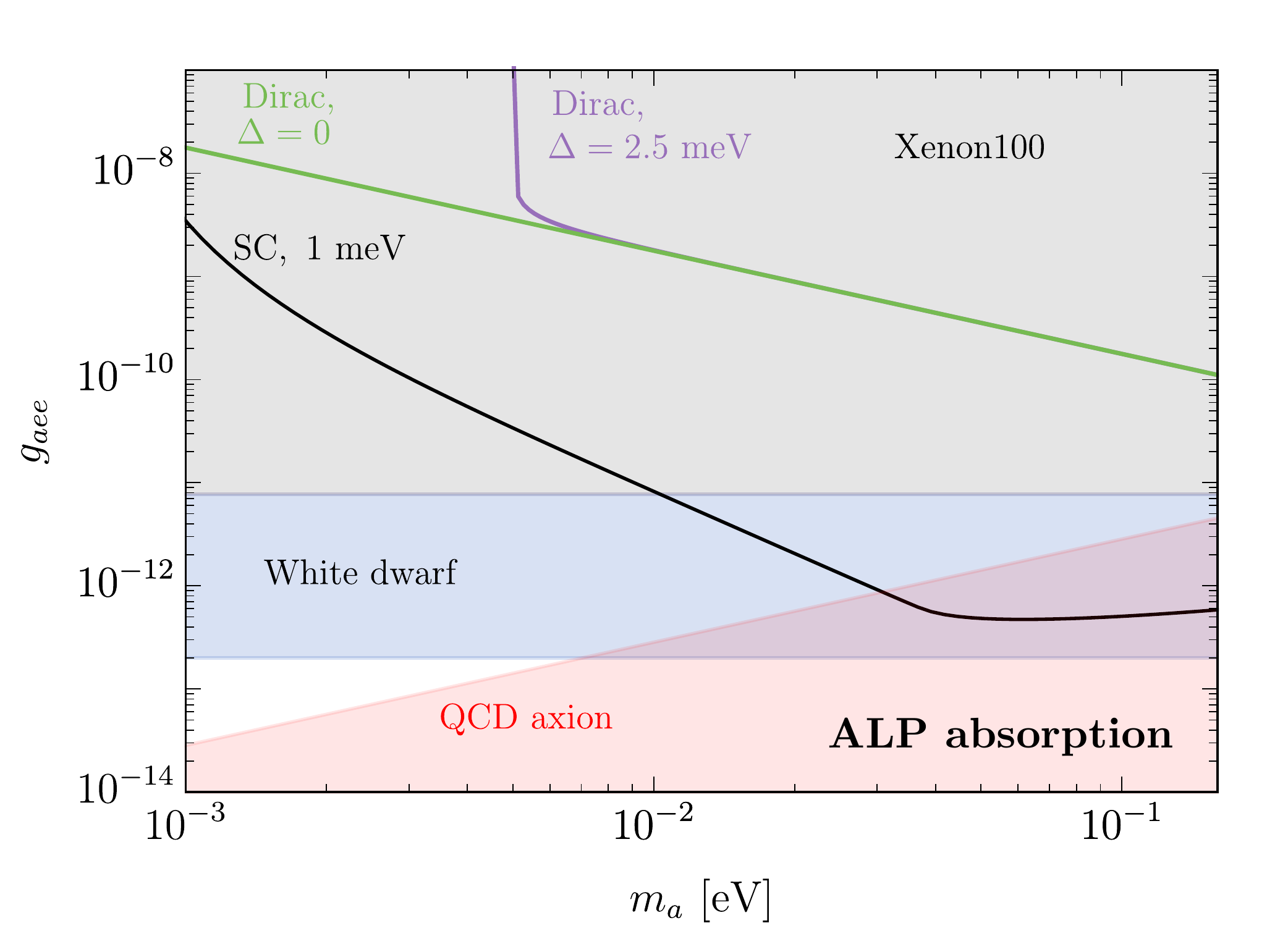}
\caption{ \label{fig:ALPabsorption}
Projected reach for absorption of axion-like particles (ALP) in Dirac materials, given in terms of the ALP-electron coupling $g_{aee}$. We show the expected background-free 95\% C.L. sensitivity (3.0 events) that can be obtained with 1 kg-yr exposure.  The green (purple) curves are gapless (gapped) isotropic Dirac materials with $\rho_T=10$~g/cm$^3$ and all other parameters as in Fig.~\ref{fig:SMkinmix}. We cut off the plot at $m_{A'} = 2 \Lambda v_F = 160 \ \meV$, the largest energy deposit consistent with the linear dispersion relation with momentum cutoff $\Lambda = 0.2 \ \keV$. We also show the reach of superconductors with a 1~meV threshold~\cite{Hochberg:2016ajh} (black) as well as constraints from Xenon100~\cite{Aprile:2014eoa} (shaded gray) and white dwarfs~\cite{Raffelt:2006cw} (shaded blue), and the QCD axion region (shaded red).
 }
\end{center}
\end{figure*}

An axion-like particle (ALP) of mass $m_a$ which comprises DM can couple to electrons via the operator
\begin{equation}
	{\cal L}\supset \frac{g_{aee}}{2 m_e} (\partial_\mu a)\bar e \gamma^\mu \gamma^5 e\,.
\end{equation}
The absorption of an ALP on electrons through this operator is related to the photon absorption rate, and is given by (see {\emph e.g.}, Refs.~\cite{Hochberg:2016ajh,Hochberg:2016sqx,Bloch:2016sjj}):
\begin{equation}
\label{eq:rateAxion}
	R_{\rm abs}^a = \frac{1}{\rho_T} \rho_\chi  \frac{3 m_a^2}{4 m_e^2}  \frac{g_{aee}^2}{e^2}\, {\rm Im}\,\epsilon_r\,.
\end{equation}
The projected reach of Dirac materials for ALP absorption is shown in Fig.~\ref{fig:ALPabsorption}, assuming the ALP comprises all of the DM, and for the same parameters as Fig.~\ref{fig:SMkinmix}. The reach of superconductors is shown for comparison~\cite{Hochberg:2016ajh}, along with the parameter space for the QCD axion  (shaded red). Constraints from Xenon100 data~\cite{Aprile:2014eoa} (shaded gray) and white dwarf cooling~\cite{Raffelt:2006cw} (shaded blue) apply, and rule out the parameter space that can be probed even by an ideal gapless Dirac material in the mass range of interest. As expected, we learn that for absorption of axion-like particles, superconductors have superior reach due to the absence of in-medium effects and larger phase space density of target electrons.


\section{Band Structure Calculations for ZrTe$_5$}
\label{app:DFT}

Among already-synthesized Dirac materials appropriate for detector targets, we identified ZrTe$_{5}$ as a strong candidate, having a linear dispersion near the Fermi level while being slightly gapped by the spin-orbit interaction.

First-principles calculations based on density functional theory (DFT) are performed using the projector augmented wave (PAW) method in the Vienna \textit{ab initio} Simulation Package
(VASP) \cite{VASP1, VASP2} code. Zr (4s, 4p, 5s, 4d), Te(5s, 5p), Se(4s, 4p), Nb(4p, 5s, 4d) and Ta(5p, 6s, 5d) electrons were treated as valence electrons, and the wavefunctions of the system were expanded in plane waves to an energy cutoff of 600 eV.  Monkhorst-Pack\cite{monkhorst1976special} $k$-point grids of 14x14x4
were used for BZ sampling. We performed calculations with the generalized gradient approximation (GGA) using the Perdew-Burke-Ernzerhof (PBE) functional \cite{PBE1}. Spin-orbit (SO) interactions are included self-consistently in all calculations. Our calculations on ZrTe$_{5}$ were performed using experimentally-determined lattice parameters and internal coordinates \cite{Fjellvaag/Kjekshus:1986}; our structural relaxations of ZrSe$_{5}$ was performed including DFT-D3 van der Waals corrections \cite{Grimme:2006}.

\begin{figure*}[t!]
\begin{center}
\includegraphics[width=\textwidth]{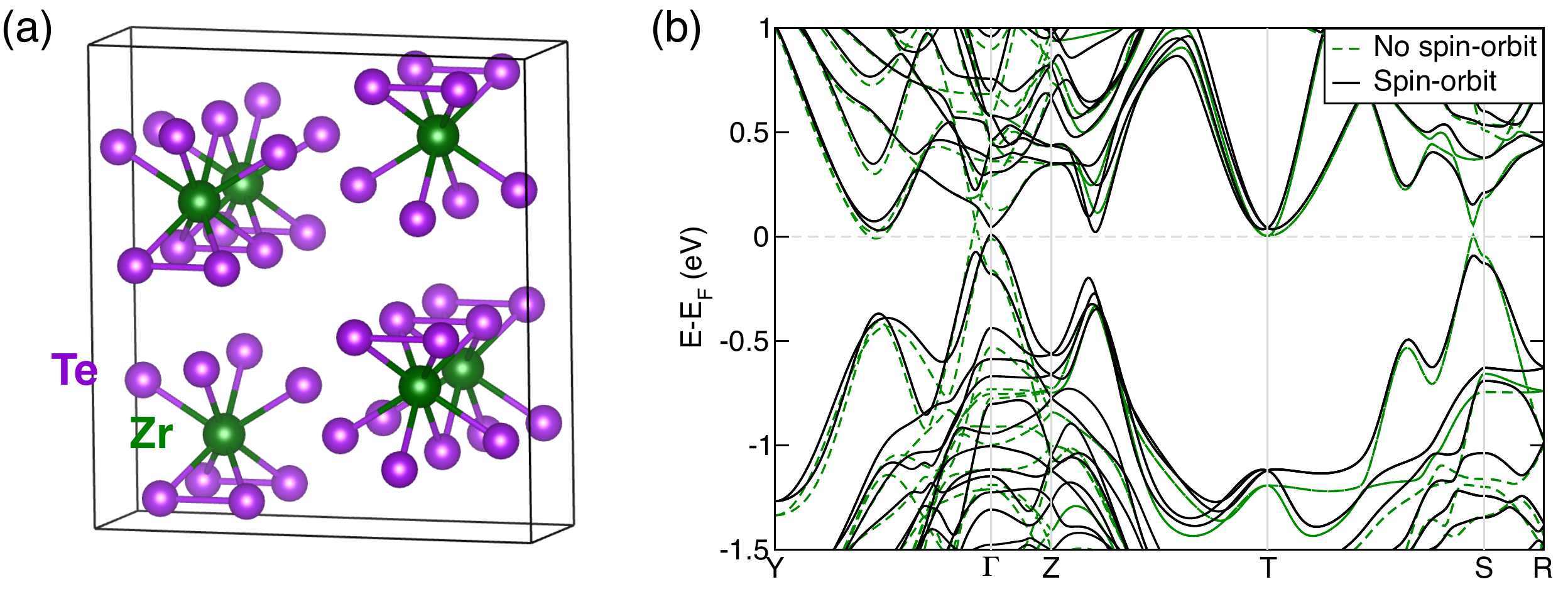}
\caption{ \label{fig:structure_bands}
(a) ZrTe$_{5}$ in the \textit{Cmcm} space group. (b) Calculated electronic band structure for ZrTe$_{5}$ with and without spin-orbit coupling. The Fermi level is set to 0 eV and marked by the dashed line.
 }
\end{center}
\end{figure*}

ZrTe$_{5}$ crystallizes in the \textit{Cmcm} structure (Space Group No. 63) as shown in Fig.~\ref{fig:structure_bands}(a). Each Zr ion is eight-fold-coordinated by Te atoms, which occupy three inequivalent lattice sites. The precise nature of the topological character of the ZrTe$_{5}$ electronic structure has been controversial, with several conflicting experiments concluding it to be a Dirac semimetal \cite{chen_magnetoinfrared_2015,SWS16,Li2016,CZS15,Zheng2016,yuan_observation_2016}, a topological insulator \cite{Chen2017,li_experimental_2016,wu_evidence_2016,manzoni_evidence_2016, nair_et_al:2017,ZWY17}, and a normal semiconductor \cite{moreschini_nature_2016}. Our first-principles PBE calculations of the electronic band structure show a Dirac cone near $\Gamma$ without spin-orbit coupling which is then slightly gapped (to 35 meV) with the inclusion of the spin-orbit interaction, consistent with previous DFT calculations~\cite{WDF14,FLC17}. We note that although DFT-GGA-SO is not expected to be qualitative for the band gap, our calculations are very consistent with previous experimental findings~\cite{wu_evidence_2016,li_experimental_2016,XSY17}. Table \ref{tab:ZrTe5} lists the material parameters we use to calculate DM scattering rates, with the theoretical values derived from the DFT calculations, and the experimental values used from the references given. If no experimental value is listed, we use the theoretical value. Our estimate of $\Lambda$ was derived from the distance between the $\Gamma$ and $Z$ points in the BZ.

\begin{table*}[t!]
\begin{centering}
\begin{tabular}{ccc} \hline
Parameter~~ & ~~value (th.) ~~ & ~~value (exp.) \\ \hline
$v_{F,1}$ & $2.9 \times 10^{-3}c~(v_{F,x})$ & $1.3 \times 10^{-3}c~(v_{F,xy})$ \cite{Zheng2016} \\
$v_{F,2}$ & $5.0 \times 10^{-4}c~(v_{F,y})$ & $6.5 \times 10^{-4}c~(v_{F,yz})$ \cite{Zheng2016}\\
$v_{F,1} $ &  $2.1 \times 10^{-3}c~(v_{F,z})$ & $1.6 \times 10^{-3}c~(v_{F,xz})$ \cite{Zheng2016}\\
$2\Delta$ (meV) & $35$ & $23.5$ \cite{XSY17} \\
$\Lambda$ (keV) & $0.14$ &\\
$g$ & 4 &\\
$\kappa_{xx}$ & $187.5$ & \\
$\kappa_{yy}$ & $9.8$ & \\
$\kappa_{zz}$ & $90.9$ &\\
$\rho_T$ (g/cm$^3$) & $6.1$  & \\
$n_e$ ($e^-$/kg) & $8.3 \times 10^{23}$ & \\
$V_{\rm uc}$ (\ang$^3$) & 795 &\\ \hline

\end{tabular}
\caption{Material parameters for ZrTe$_5$. $v_{F,i}$ ($i = 1,2,3$) are Fermi velocities, $2\Delta$ is the gap, $\Lambda$ is the linear dispersion cutoff, $g = g_s g_C$ is the product of spin and Dirac cone degeneracies, $\kappa_{ii}$ ($i = 1,2,3$) are principal components of the background dielectric tensor, $\rho_T$ is the density, $n_e$ is the mass density of Dirac valence-band electrons, and $V_{\rm uc}$ is the unit cell volume. Where no experimental value is listed, we use the theoretical value. The theoretical values of the Fermi velocities were calculated along the high-symmetry directions, while the experimental values are mid-plane velocities. For the experimental value of $2\Delta$, we take the median of the range of values presented in \cite{XSY17}. $\Lambda$ was taken to be the distance between the $\Gamma$ and $Z$ points in the BZ, see Figs.~\ref{fig:structure_bands} and \ref{allbands}. The static ion-clamped dielectric tensor $\kappa_{ij}$ was calculated using density functional perturbation theory. The unit cell is defined as containing 4 formula units, see Fig.~\ref{fig:structure_bands}(a).}
\label{tab:ZrTe5}
\end{centering}
\end{table*}

\begin{figure*}[t!]
\begin{center}
\includegraphics[width=\textwidth]{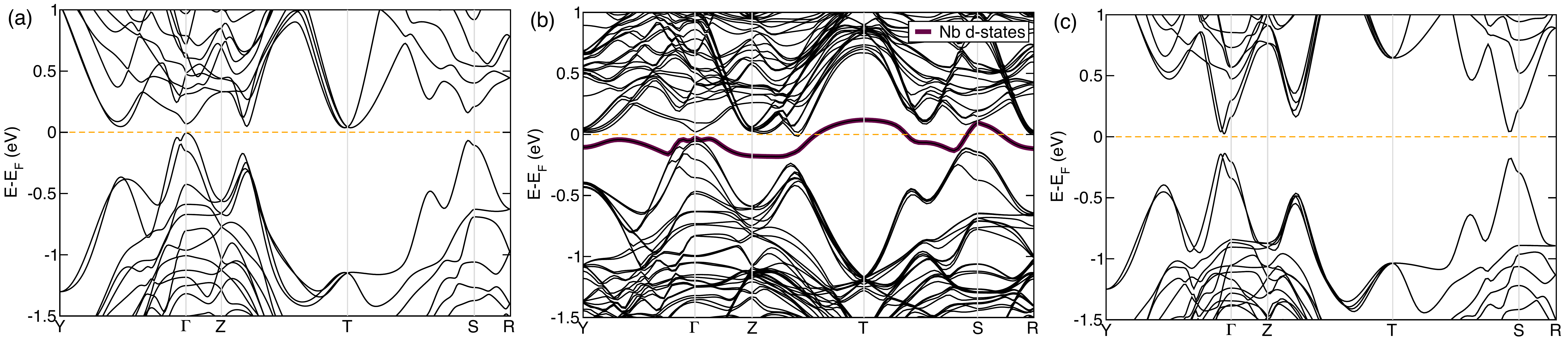}
\caption{ \label{allbands}
Calculated band structure for (a) stoichiometric ZrTe$_{5}$ with 99\% the lattice volume of the experimental lattice parameters, (b) ZrTe$_{5}$ with 12.5\% Nb substitution on the Zr site and (c) ZrSe$_{5}$. In (b), the Nb d-states near the Fermi level are indicated by the weighted line. In each plot the Fermi level is set to 0 eV and marked by the dashed line.
 }
\end{center}
\end{figure*}

While the band structure shows the gapping of the Dirac cone near $\Gamma$, the Fermi level cuts the top of the band to form a hole-like pocket. To engineer a semiconducting band structure, with the Fermi level in the gap, we recompute the band structure of electron-doped ZrTe$_{5}$ by adding a small fraction of electrons per unit cell and compensating this additional electron density with a uniform positive background. We find that electron doping by 0.2 electrons per unit cell shifts the Fermi level into the gap. Alternatively, Fig.~\ref{allbands}(a) shows the band structure for stoichiometric ZrTe$_{5}$ at 99\% of the experimental lattice volume. We find that a small amount of pressure results in the desired band structure with the Fermi level now in the gap. This could potentially be achieved experimentally by epitaxial growth on a substrate with a slightly smaller in-plane lattice parameter or by chemical substitution of ions with a smaller radius.

We next consider chemical substitution. Since the ZrTe$_{5}$ bands near the Fermi level consist primarily of Te-p states, we consider substitution on the Zr site by Nb and Ta. We calculate the band structure of substitution of one Nb/Ta for eight formula units, resulting in electron doping of 0.25 electrons per formula unit as shown in Fig.~\ref{allbands}(b) for the Nb case. While the Fermi level shifts as expected, Nb contributes d-states near the Fermi level, making the material a metal. The same also occurs for the case of Ta substitution. Substitution of Te with Br alters the band structure near the Fermi level as well.

In light of this, and with the additional motivation of reducing the band gap, we consider replacing Te with Se in the hypothetical new compound ZrSe$_{5}$ in the same \textit{Cmcm} structure as shown in Fig.~\ref{allbands}(c). This chemical substitution has three effects on the electronic properties of the material. Firstly, the smaller ionic radius of Se reduces the total volume of the compound which results in a Fermi level in the gap without any external pressure; however, this also has the undesired effect of increasing the band gap. Independent of the volume change, the lower spin-orbit coupling in Se reduces the spin-orbit splitting of the  bands to $2\Delta \simeq 15 \ \meV$. Therefore, our DFT estimates suggest that ZrTe$_{5}$ with a small amount of Se alloying could provide a more desirable volume contraction and spin-orbit-driven reduction in band gap. Interestingly, another Dirac cone is present in the ZrSe$_{5}$ compound, which doubles the number of Dirac cones and Dirac valence-band electrons per unit cell. Since the DM scattering rate scales as $n_e/g$, from stoichiometry alone we would expect the overall rate to increase by a factor of $m_{\rm Te}/m_{\rm Se} \simeq$ 1.5 for ZrSe$_{5}$, with additional increases near threshold from the reduced gap. Neither ZrSe$_{5}$ nor Zr(Te,Se)$_{5}$ have yet been synthesized; should synthesis be possible, these compounds may be  be promising targets for DM detection.

\newpage

\bibliographystyle{JHEP}
\bibliography{LDMMaterialsBib}

\end{spacing}

\end{document}